\documentclass{aa}  
\usepackage{dblfloatfix}
\usepackage{graphicx}
\usepackage{txfonts}
\usepackage{amsmath}
\newcommand{\m}[1]{\mathrm{#1}}
\usepackage{mathrsfs,amsmath} 
\usepackage[
  colorlinks=true,
  linkcolor=blue,
  citecolor=blue,
  urlcolor=blue
]{hyperref}

\usepackage{lipsum}
\usepackage{subcaption}         
\usepackage{lscape}             
\usepackage{lineno}             
\usepackage{placeins}           
\usepackage{natbib}
\bibpunct{(}{)}{;}{a}{}{,} 
\usepackage{hyperref}
\usepackage{ulem}

\usepackage[dvipsnames]{xcolor} 

\newcommand{\kim}[1]{\textcolor{Black}{#1}}

\defcitealias{mcnallyMigratingSuperEarthsLowviscosity2019}{Mc19}
\defcitealias{wafflard-fernandezIntermittentPlanetMigration2020}{WB20}
\defcitealias{meinersPlanetMigrationALMA}{M26}
\defcitealias{ziamprasHaltingMigrationSuperEarths2025}{Z25}

\newcommand{\Zmigration}[1]{\citetalias{ziamprasHaltingMigrationSuperEarths2025}}

\begin{document}

\title{Formation of multiple dust rings and gaps in protoplanetary discs by a single migrating planet}
\subtitle{\kim{Radiative discs and observational signatures}}

   \author{K. M. Weiskopf\inst{\ref{LMU}}, S. C. Meiners\inst{\ref{ZAH}},  A. Ziampras\inst{\ref{LMU},\ref{MPIA}} and C. P. Dullemond\inst{\ref{ZAH}}
        }

   \institute{Ludwig-Maximilians-Universität München, Universitäts-Sternwarte, Scheinerstr. 1, 81679 München, Germany\label{LMU}
   \and Institute for Theoretical Astrophysics, centre for Astronomie (ZAH)\label{ZAH}, Heidelberg University, Albert Ueberle Str. 2, 69120 Heidelberg,
   \and Max Planck Institute for Astronomy, K{\"o}nigstuhl 17, 69117 Heidelberg, Germany\label{MPIA}
Germany}

   \date{Received XX, 20XX}

  \abstract 
   {Dust structures in protoplanetary discs have been widely observed, and their creation remains an active field of research. Several possible origins have already been explored, including magnetohydrodynamic processes, shadows, and planet--disc interactions.}
   {The goal of this paper is to investigate whether a single migrating planet in a low-viscosity disc, including radiative processes, is capable of generating observable dust structures. We examine both the lifetime of such structures and potential asymmetries within them.} 
   {We performed a set of high-resolution, two-dimensional hydrodynamic simulations of migrating planets using three different equations of state: isothermal, constant $\beta$-cooling, and an adaptive $\beta$ model. We included dust in all simulations and post-processed the resulting dust density profiles  to create radiative transfer images.}
   {For all equations of state considered, the planet undergoes one or several migration jumps, each producing dust rings and gaps. The lifetime of these structures depends on the phase of slow migration preceding and occurring between jumps, but in all cases they remain visible for at least 400\,kyr. We find that cooling has a decisive effect on the migration behaviour and the number of jumps, but no measurable influence on the lifetime of the dust structures. The structures exhibit relatively few asymmetries and large-scale vortices persist for an average of only 90 \,kyr.}
   {Our models highlight the ability of planets to open multiple gaps while migrating and stress the importance of a realistic cooling model. Care should be taken when interpreting and comparing such models directly with observations.}


\authorrunning{K. M. Weiskopf, S. C. Meiners, A. Ziampras and C. P. Dullemond}
\titlerunning{\kim{Radiative discs and observational signatures}}
\maketitle

\section{Introduction}

The Disk Substructures at High Angular Resolution Project (DSHARP) survey (\citealt{andrewsDiskSubstructuresHigh2018}) observed the $\lambda = 1.25$\,mm continuum and ${ }^{12} \mathrm{CO} \mathrm{~J}=2-1$ line emission of 20 nearby protoplanetary discs. They confirm without doubt that the brightest and largest protoplanetary discs can contain visible dust rings and gaps, as well as asymmetries in their dust profile. At $\lambda \approx 1.25$\,mm, the emission is dominated by thermal radiation from millimetre-sized dust grains in the disc mid-plane, which allows identification of disc substructures. Similar observations have also been reported by the Ophiuchus DIsk Survey Employing ALMA (ODISEA) (\citealt{orcajoOphiuchusDIskSurvey2025}), exoALMA (\citealt{teagueExoALMAScienceGoals2025}), the Molecules with ALMA at Planet-forming Scales (MAPS) (\citealt{Oeberg_et_al_2021}), and the ALMA survey of the Lupus (\citealt{Ansdell_et_al_2018}) and Taurus (\citealt{Long_et_al_2018}) star-forming regions. A better understanding of the formation of these dust structures would provide a deeper insight into the dynamics of the disc.

Several theoretical formation mechanisms to explain the development of substructure have been identified. These mechanisms include the formation of pressure bumps at the inner edge of the dead zone \citep{armitage-etal-2001,varniereRevivingDeadZones2006,flockGapsRingsNonaxisymmetric2015,flock-etal-2017a,roberts-etal-2025}, spontaneous dust ring formation due to the streaming instability \citep{youdin-goodman-2025,ostertag-flock-2025}, differential radial drift \citep{jiang-ormel-2023}, magnetic effects \citep{bethune-etal-2016,riols-etal-2020}, thermal instabilities \citep{ueda-etal-2021,sudarshan-etal-2026,owen-2020}, warps \citep{smallwood-etal-2024a,smallwood-etal-2024b}, and shadow-driven structures \citep{su-bai-2024,qian-wu-2024, ziamprasSpiralsRingsVortices2025,zhang-etal-2025}.

Another commonly proposed explanation is the presence of embedded planets. They launch spiral waves due to a resonance between the planet's gravitational potential and the epicyclic motion of the disc material (\citealt{goldreichExcitationDensityWaves1979, ogilvieWakeGeneratedPlanet2002}). For sufficiently high planetary masses or low disc viscosities, the excited spiral waves steepen into shocks, clearing material from the planet’s vicinity and thus opening a gap (\citealt{goodmanPlanetaryTorquesViscosity2001,rafikovPlanetMigrationGap2002}). \citet{zhangDiskSubstructuresHigh2018, mirandaGapsRingsProtoplanetary2020, zhangEffectsDiscSelfgravity2020}, and \citet{ziamprasImportanceRadiativeEffects2020} show that a single non-migrating planet can open multiple gaps and study the effects of radiative processes on gap formation. 

\cite{mcnallyMigratingSuperEarthsLowviscosity2019} (hereafter \citetalias{mcnallyMigratingSuperEarthsLowviscosity2019}) and later \cite{wafflard-fernandezIntermittentPlanetMigration2020} (hereafter \citetalias{wafflard-fernandezIntermittentPlanetMigration2020}), who considered higher viscosities, investigated the migration of a single planet in an isothermal disc and discovered migration jumps, referred to as `intermittent migration'. These jumps can create dust ring and gap structures, which remain visible when plotting the continuum emission at (sub)millimetre wavelengths. \cite{meinersPlanetMigrationALMA} (hereafter \citetalias{meinersPlanetMigrationALMA}) continued the analysis of migration, covering different aspect ratios and viscosities, and disabling the gas self-gravity in all simulations. They also observed intermittent migration jumps, further confirming their existence.

In general, planetary migration is highly dependent on the level of turbulence within the disc, typically modelled as an effective `viscosity' with an $\alpha$ parameter \citep{shakuraBlackHolesBinary1973}. For $\alpha \gtrsim 10^{-3}$, planets migrate in the classical type-I and type-II regimes (\citealt{paardekooper-etal-2011,kleyPlanetDiskInteractionOrbital2012}), depending on whether they can open a gap. Type-I migration is driven by an interplay of Lindblad and corotation torques, resulting in rapid migration. For an Earth-sized planet, this corresponds to a migration timescale of $\tau_{mig} \sim 10^{5}$\,yr (\citealt{kleyPlanetDiskInteractionOrbital2012}).

If the planet opens a sufficiently deep gap and clears its corotation region, only very few Lindblad modes act, which produces a negligible net torque. \cite{kleyPlanetDiskInteractionOrbital2012} suggest that the planet is now coupled with the viscous evolution of the disc, as the deep gap prevents material from entering the inner disc. On the other hand, \citet{Duermann-Kley-2015} found that the planet only prevents gas from crossing the gap in its immediate vicinity, challenging the idea that the planet is coupled to the viscous evolution. This results in migration rates that differ significantly from the viscous timescale. Furthermore, \citet{scardoniInwardOutwardMigration2022} show that a planet in the type-II regime might switch from inward to outward migration, decoupled from the viscous evolution, but still migrating with a similar rate. Additionally, a planet might undergo phases of very rapid migration, which are often labelled as type-III migration (\citealt{massetRunawayMigrationFormation2003, rometschMigrationJumpsPlanets2020}).

In recent years, a number of studies have suggested that the disc's viscosity takes lower values, with a median around $\alpha \sim 10^{-4}$ (\citealt{villenaveHighlySettledDisk2022, dullemondDiskSubstructuresHigh2018, rosottiEmpiricalConstraintsTurbulence2023}), but its value remains a topic of discussion. We further acknowledge that inferring viscosity values from observations comes with limitations. \citet{ziamprasHaltingMigrationSuperEarths2025} (hereafter \Zmigration{}) introduced new migration regimes in these nearly inviscid discs, which emerge for planets in a certain mass range. Hence, migration is more diverse than previously thought. 

As a continuation of \citetalias{mcnallyMigratingSuperEarthsLowviscosity2019} and \citetalias{meinersPlanetMigrationALMA} work, the goal of this paper is to investigate the ability of a single planet to open multiple gaps while including radiative processes on the disc. We tailor our setups to model the outer disc, ensuring comparability with ALMA observations. We further analyse the lifetime and appearance of the created rings and gaps and assess the observability of these structures.

In Sect.~\ref{sec:physics and numerics} we introduce our physical model and numerical framework. We present our results in Sect.~\ref{sec:results} and discuss them in Sect.~\ref{sec:discussion}. We present our conclusions in Sect.~\ref{sec:conclusion}.

\section{Physics and numerics}\label{sec:physics and numerics}

In this section, we briefly describe the physical model and numerical setup. We adopted our radiative cooling model from \citet{ziamprasModellingPlanetinducedGaps2023} and summarise it below.

\subsection{Physical model}\label{sec:physical background}

We modelled a vertically integrated disc around a central star. The gas is assumed to be ideal with a surface density $\Sigma$, internal energy $e$, and velocity $\vec{u}$. Its equations of motion are given by the Navier--Stokes equations in cylindrical coordinates $\{ R, \phi \}$ with the adiabatic index $\gamma$: 

\begin{align}\label{Navier stokes}
    &\frac{\partial \Sigma}{\partial t} + \vec{u} \cdot\, \nabla\, \Sigma = - \Sigma \nabla \cdot \vec{u},\\
    &\Sigma \frac{\partial \vec{u}}{\partial t} + \Sigma (\vec{u} \cdot \nabla) \vec{u} = -\nabla P - \Sigma \vec{g} + \nabla \cdot\bar{\sigma},\\
    &\frac{\partial \Sigma e}{\partial t} + \nabla \cdot (\Sigma e \vec{u})  = - P \nabla\cdot\vec{u} -\Sigma c_{\mathrm{v}} \frac{T-T_0}{\beta}\Omega_{\m{K}}. \label{energy eq}
\end{align}

Here, $\bar{\sigma}$ denotes the viscous stress tensor and $P = (\gamma -1)\Sigma e$ denotes the vertically integrated pressure. The isothermal sound speed is given by $c_{\m{s}}= \sqrt{P/\Sigma}= \sqrt{\mathcal{R} T/\mu}$, with the mean molecular mass $\mu = 2.35$. We define a pressure scale height $H=c_{\m{s}}/\Omega_{\m{K}} $, with a Keplerian frequency $\Omega_{\m{K}}=\sqrt{\text{G}M_{*}/R^{3}}$. The aspect ratio $h$ is given by $h = H/R$, where $R$ is the radial distance to the central star. The viscosity follows the model from \citet{shakuraBlackHolesBinary1973}, with $\nu = \alpha c_{\m{s}} H$. We define the vorticity $\omega = \vec{\nabla} \times \vec{u}$ and the inverse vortensity ($\mathcal{IV}$) of the gas as
\begin{equation}
    \mathcal{IV} = \frac{\Sigma}{\omega}\,.
\end{equation}
We used this quantity to confirm the existence of small vortices in the disc, which present as local maxima. We used the inverse vortensity, since it scales with the coorbital mass deficit, which is related to intermittent migration, scales with $\mathcal{IV}$ \citep{massetRunawayMigrationFormation2003,wafflard-fernandezIntermittentPlanetMigration2020}. 

\begin{figure}[t]
   \centering
   \includegraphics[width=\hsize]{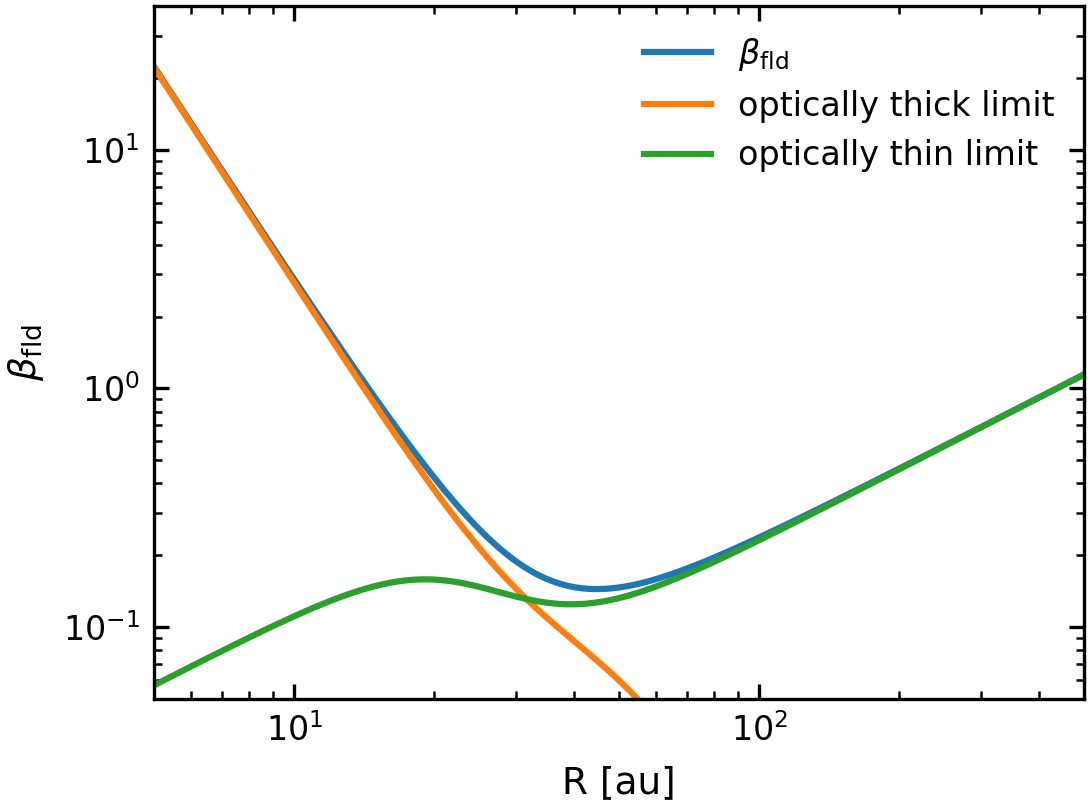}
      \caption{Dimensionless cooling timescale $\beta_{\m{fld}}$ across the
extend of the disc at $t= 0$\,kyr. We also plot the approximations for the optically thick and optically thin limit.}
         \label{beta fld profile}
\end{figure}

We did not include viscous heating because the paper focuses on planets predominately in the outer regions of the disc. Following \citet{ziamprasModellingPlanetinducedGaps2023} and \citet{mirandaGapsRingsProtoplanetary2020}, we defined cooling timescales for radiative loss from the surface (hereafter surface cooling), cooling of the mid-plane through radial radiative diffusion (now mid-plane cooling), and flux-limited diffusion (FLD). Estimating a cooling timescale using $t_{\m{cool}} \sim e/\dot{e}$ (\citealt{ziamprasImportanceRadiativeEffects2020}) results in a dimensionless cooling timescale for surface cooling,
\begin{equation}\label{beta surf}
    \beta_{\m{surf}} = \frac{e}{\vert Q_{\m{cool}} \vert }\, \Omega_{\m{K}}, \quad Q_{\m{cool}} = - 2 \sigma_{\m{SB}} \frac{T^{4}}{\tau_{\m{eff}}}\quad\tau_{\m{eff}} = \frac{3 \tau}{8} + \frac{\sqrt{3}}{4} + \frac{1}{4\, \tau},
\end{equation}

where $\sigma_{\m{SB}}$ denotes the Boltzmann constant and the effective optical depth $\tau_\text{eff}$ follows \citet{hubenyVerticalStructureAccretion1990}. Defining a timescale for mid-plane cooling is more difficult since the diffusion length $l_{\m{d}}$ is not constant. \citealt{mirandaGapsRingsProtoplanetary2020} show that if we thermally perturb the disc, we can approximate the $m$-th azimuthal Fourier component of this perturbation. Combining this approximation with the assumption that the dominant diffusion length $l_{\m{d}}$ is on the order of $H$ (\citealt{ziamprasModellingPlanetinducedGaps2023}) results in
\begin{equation}\label{beta mid}
    \beta_{\m{mid}} = \frac{\Omega_{\m{K}}}{\eta} \left( H^{2} + \frac{1}{3} l_{\m{rad}}^{2} \right). 
\end{equation}
We define $l_{\m{rad}} = 1/(\kappa \rho)$, with the opacity given by $\kappa$. Combining Eq.~\ref{beta surf} and Eq.~\ref{beta mid} produces a $\beta$-cooling description of FLD:
\begin{equation}\label{beta FLD}
    \beta_{\m{fld}}=\frac{\beta_{\m{surf}}}{1+f}, \quad f=\frac{\beta_{\m{surf}}}{\beta_{\m{mid}}}=16 \pi \frac{\tau_{\m{eff}}}{\tau_{\m{R}}} \frac{\tau_{\m{P}}^{2}}{6 \tau_{\m{P}}^2+\pi}.
\end{equation}
We assume that $\tau_{\m{R}} = \tau_{\m{P}}$, where $\tau_{\m{R}}$ is the Rosseland mean optical depth and $\tau_{\m{P}}$ is the Planck mean optical depth. In contrast to a full radiative description, $\beta_{\mathrm{fld}}$ cools the disc only locally. \citet{ziamprasModellingPlanetinducedGaps2023} show that $\beta_{\mathrm{fld}}$ can substitute full FLD with respect to the cooling time in shock-driven gap opening, but it cannot reproduce all migration-related effects of FLD (\citealt{ziamprasMigrationLowmassPlanets2024}). With this, we define three different equations of state, adjusting the complexity of them step-wise: 

\begin{enumerate}
\setlength{\itemsep}{6pt}
    \item Locally isothermal: there is no compressive heating and the disc is in thermal equilibrium. The temperature is given by a balance between irradiation and surface cooling following \citet{menouLowMassProtoplanetMigration2004}:
    
    \begin{equation}
        \label{eq:temperature-profile}
        T(R)=18.07\,\text{K}\,\left( \frac{R}{\text{50\,au}} \right)^{-3/7}\, \Rightarrow \,\,  h(R) = 0.06\, \left( \frac{R}{\text{50\,au}} \right)^{2/7}.
    \end{equation}
    
    With $c_{\m{s}} \propto \sqrt{T}$, the pressure is directly linked to the surface density.

    \item Constant $\beta=1$: the system exponentially cools back to the steady-state temperature profile given by Eq.~\ref{eq:temperature-profile} and is only heated by shocks, as in Eq.~\eqref{energy eq}. 

    \item Adaptive $\beta$: the cooling timescale instead follows a density-, temperature-, and opacity-dependent formulation through Eqs.~\eqref{beta surf}--\eqref{beta FLD}. This represents an accurate model of $\beta$ following \citet{mirandaGapsRingsProtoplanetary2020,ziamprasModellingPlanetinducedGaps2023,ziampras-etal-2026}. Figure~\ref{beta fld profile} shows $\beta_{\m{fld}}$ over the disc extent for our setup at $t = 0$\,kyr.
\end{enumerate}

We set $\beta = 1$ for the constant $\beta$-cooling simulation, since it translates to strong radiative damping of planetary wakes and therefore to a single-gap configuration \citep{mirandaPlanetDiskInteraction2020, ziamprasModellingPlanetinducedGaps2023,zhangDependenceStructurePlanetopened2024}. This minimises the risk of secondary gaps produced by the planet \citep{zhangDiskSubstructuresHigh2018} affecting its inward migration track, but does not necessarily eliminate it \citep[e.g.][]{lega-etal-2021,wafflard-fernandez-lesur-2025}.

As shown by \cite{ziamprasMigrationLowmassPlanets2024}, radiative processes can modify the local vortensity $\varpi$ through baroclinic forcing $\mathcal{S}$:
\begin{equation}\label{baroclinic forcing}
    \frac{\partial \varpi}{ \partial t } + (\vec{v} \cdot \nabla)\, \varpi = \frac{\nabla \Sigma \times \nabla P }{\Sigma^{3}} = \frac{P}{T R\Sigma^3}\left(\frac{\partial \Sigma}{\partial R} \frac{\partial T}{\partial \phi}-\frac{\partial \Sigma}{\partial \phi} \frac{\partial T}{\partial R}\right) = \mathcal{S}.
\end{equation}

Specifically, a parcel of gas approaching the planet within the horseshoe region gains (loses) vortensity as it U-turns behind (ahead of) the planet. Cooling results in an asymmetry between the two sides and typically leads to vortensity growth within the planet's horseshoe region, with the effect maximised for $\beta\sim1$. This can substantially influence the planet's migration, as we discuss in the following. We note that using $\beta_\text{mid}$ as a substitute for in-plane diffusion can exaggerate this effect \citep{ziamprasMigrationLowmassPlanets2024}, however, we chose to use this approach due to the prohibitive computational costs of a fully radiative model.

\subsection{Numerical setup}\label{sec:numerical setup}

We used the hydrodynamic code \texttt{FARGOCPT} (\citealt{rometschFARGOCPT2DMultiphysics2024}), a successor of \texttt{Fargo} (\citealt{massetFARGOFastEulerian2000}), to create a vertically integrated, cylindrically symmetric ($R,\phi$) disc extending from 5--500\,au. We analysed the data using \texttt{python}. We set the grids resolution to 16 cells per scale height at 50\,au in radial and azimuthal directions. This corresponds to $1230 \times 1676$ approximately square cells, logarithmically spaced in the radial direction. The inner and outer boundaries are set to outflow, both with damping zones in front of them. These damping zones range from 5--6\,au and $425$--$500$ au, respectively, damping the values exponentially on a timescale of $10^{-1}\, 2\pi/\Omega_{\m{K}}$ evaluated at the boundary. We tested different configurations of boundary conditions, damping zones, and exponential cut-off beforehand. The chosen setup prevented any pile-ups at the inner boundary and reflections of density waves at the outer boundary. We set the initial surface density to

\begin{equation}
    \Sigma = 10\,\frac{\m{g}}{\m{cm}^{2}} \cdot \left( \frac{R}{50\, \m{au}} \right)^{-1} \, C_{\m{o}}(r).
\end{equation} 
 An outer surface-density profile cut-off is given by $C_{\m{o}}(r)$, implemented starting at 350\,au:
\begin{equation}
C_{\mathrm{o}}=\left(1+\exp \left(\frac{( - (r - r_{\mathrm{oC}} )}{w_{\mathrm{oC}}}\right)\right)^{-1}.
\end{equation}
Here, $r_{\mathrm{oC}} = 350$ au is the cut-off point and $w_{\mathrm{oC}} = 35$ au the cut-off width. The total disc mass between 5--500\,au is $\approx 0.12 \, \m{M}_{\odot} $.
We disabled gas self-gravity, since the Toomre parameter $Q = \frac{c_s \Omega_{\mathrm{K}}}{\pi \mathrm{G} \Sigma} > 1$ over the entire disc. We used a turbulent $\alpha$-value of $\alpha =10^{-4}$ for our viscosity. The disc aspect ratio follows Eq.~\eqref{eq:temperature-profile}. The central star has solar parameters, and the grid frame is centred on the star. All disc parameters fall within the range characteristic of DSHARP discs, except for the radial extent (\citealt{dullemondDiskSubstructuresHigh2018,huangDiskSubstructuresHigh2018, zhangALMASurveyGas2025}). We set the planet mass $M_{\m{p}} = 100\, \text{M}_{\oplus}$. We define the thermal mass as
\begin{equation}
   \label{eq:thermal-mass}
   M_{\m{th}} = 2 \frac{h_{\m{p}}^{3}}{3}M_{*},
\end{equation}
and the feedback mass as
\begin{equation}
\label{eq:feedback-mass}
M_{\mathrm{f}} \approx 3.8\left(\frac{Q}{h}\right)^{-5 / 13} M_{\mathrm{th}}.
\end{equation}
If $M_{\m{p}} \approx M_{\m{th}} $, the planetary wakes steepen into shock waves immediately after launch, while a planet with $M_{\m{p}} \approx M_{\m{f}} $ has the minimal mass required to significantly modify its surroundings. \citetalias{mcnallyMigratingSuperEarthsLowviscosity2019} and \Zmigration{} define different migration regimes based on the ratios of $M_{\m{p}}$ to $M_{\m{th}}$ and $M_{\m{p}}$ to $M_{\m{f}}$. In this paper $M_{\m{p}} \approx 2\, M_{\m{th}}$ and $\approx 3.4\, M_{\m{f}}$. We placed the planet at $R_{\m{p}}=50$\,au and allowed it to migrate immediately. The planet's potential is given by
\begin{equation}
    \Phi_{\m{p,2D}} = - \frac{\text{G}M_{\m{p}}}{\sqrt{d^{2} + s^{2}}}, \quad \vec{d} = \vec{R} - \vec{R}_{\m{p}}\,,
\end{equation}
with a smoothing length $s = \epsilon H$ and a smoothing parameter $\epsilon = 0.6$to account for the vertical stratification of the disc (\citealt{mullerCircumstellarDisksBinary2012}). We let the system evolve for $530$\,kyr, corresponding to 1500 orbits at $R = 50$\,au.

The planet and star influence each other's motion. Because we fixed our frame onto the star, the frame is non-inertial, introducing additional forces. In \texttt{FARGOCPT}, these forces are accounted for by the so-called indirect terms. We include the indirect acceleration of the star due to the star--planet system orbiting their common centre of mass. However, because we did not include self-gravity in our simulations, we disabled the disc-disc indirect term `ITdd' described in \citet{Crida-et-al-2025} (i.e. the back-reaction of the disc's acceleration on the star onto the disc), which would otherwise artificially enhance vortices and potentially trigger the reflex instability (\citealt{Crida-reflex-2025}). 

We also included dust in our simulation. In \texttt{FARGOCPT}, this is achieved using Lagrangian super-particles with a particle density of $2.08\,\m{g}/\m{cm}^{3}$ following the DIANA standard model. They are split into three sets of 33\,000 particles with sizes $a_\mathrm{d}$ fixed at $3\,\mu$m, $30\,\mu$m, and $0.3$\,mm. These sizes correspond to Stokes numbers of $\m{St} = 10^{-4},\, 10^{-3}$, and $10^{-2}$ at 50\,au, respectively. Their surface density profile follows the gas surface density without the cut-off. The initial dust-to-gas ratio is 0.01, and the total dust mass is therefore $0.0012\, \m{M}_{\odot}$, initialised with an MRN (\citep{1977MRN}) size distribution such that $\Sigma_{\mathrm{d},i}\propto a_{\mathrm{d},i}^{0.5}$ \citep{mathis-etal-1977}. We did not include dust feedback onto the gas.\\

Table ~\ref{tab:sim-list} lists the simulations. We analyse the migration of some simulations in detail in Sect.~\ref{migration models} and discuss their post-processing in Sect.~\ref{post processing}.

\renewcommand{\arraystretch}{1.3}

\begin{table}[t]
\caption{Overview of the simulations considered in this study.}
    \centering
    \begin{tabular}{c|c|c} 
    Name & EOS & $M_{\mathrm{p}}$ [$\mathrm{M}_{\oplus}$] \\
        \hline
         \texttt{iso}& isothermal & 100 \\
         \texttt{beta-01}& $\beta = 0.01$ & 100 \\
         \texttt{beta}& $\beta = 1$ & 100 \\
         \texttt{beta-3}& $\beta = 3$ & 100 \\
         \texttt{beta-7.5}& $\beta = 7.5$ & 100 \\
         \texttt{adapt}& $\beta = \beta_{\m{fld}}$ & 100 \\
         \texttt{adapt-low-mass}& $\beta = \beta_{\m{fld}}$ & 62 
         \\
    \end{tabular}
    \label{tab:sim-list}
\end{table}

\renewcommand{\arraystretch}{1}

\section{Results}\label{sec:results}

\subsection{Migration regimes}\label{sec:migration regimes}

One aspect of the simulation analysis is the identification of the planet's migration regimes and their link to the feedback and thermal masses $M_\text{f},\, M_\text{th}$ (see Eqs.~\eqref{eq:thermal-mass}~and ~\eqref{eq:feedback-mass}). \citetalias{mcnallyMigratingSuperEarthsLowviscosity2019} and \Zmigration{} demonstrated that in nearly inviscid discs, planets with masses $M_\text{f}\lesssim M_\text{p}\lesssim M_\text{th}$ migrate in the turbulent vortex-assisted regime.

The planet opens a partial asymmetric gap that is deeper behind the planet than ahead of it, resulting in a pronounced surface-density gradient at the outer gap edge \citepalias[see also Fig.~5 in][]{ziamprasHaltingMigrationSuperEarths2025}. This configuration is eventually unstable to the Rossby wave instability (RWI; \citealt{lovelaceRossbyWaveInstability1999}) and is further subject to baroclinic forcing via radiative processes that can significantly modify the vortensity around the planet's orbit \citep{ziamprasMigrationLowmassPlanets2024,ziamprasHaltingMigrationSuperEarths2025}. The change in vortensity is directly linked to a change in the surface-density profile, which produces additional RWI vortices. These RWI vortices, whether driven by baroclinic forcing or not, at the outer gap edge refill the gap and thus sustain the outer Lindblad torques (\citealt{armitageProtoplanetaryDisksPlanet2019}), driving inward migration. With a width of $\sim6$--$23$\% of the half-width of the horseshoe region, these small vortices are often only visible when inspecting the inverse vortensity ($\mathcal{IV}$) profile at high resolution. They can also be identified by plotting the relative deviation from the azimuthally averaged inverse vortensity, $\mathcal{IV}/\overline{\mathcal{IV}} -1$. We do not distinguish between two vortex-assisted sub-regimes, one without cooling and one with cooling, but note that when cooling is active, baroclinic forcing significantly alters the regime.

As $M_\text{th}$ decreases while the planet migrates inwards, the planet mass eventually becomes super-thermal and transitions into the type-II regime, ultimately stalling its migration. 

In addition, we observe intermittent migration, first introduced by \citetalias{mcnallyMigratingSuperEarthsLowviscosity2019} \citepalias[see also][]{wafflard-fernandezIntermittentPlanetMigration2020, ziamprasHaltingMigrationSuperEarths2025, meinersPlanetMigrationALMA}. Intermittent migration can be described as brief episodes of type-III migration (\citealt{massetRunawayMigrationFormation2003}), in which the dynamical co-rotation on the planet \citep{paardekooper-2014},

\begin{equation}
    \label{eq:DCT}
     \Gamma_{\text{h}}  = 2\pi\left(1-\frac{\mathcal{IV}_\text{h}}{\mathcal{IV}_\text{p}}\right)\Sigma_\mathrm{p} R_\text{p}^2 x_\text{h}\Omega_\mathrm{p}\left(\frac{\text{d}R_\text{p}}{\text{d}t} - u_R\right),
\end{equation}

becomes momentarily strongly negative due to a sharp contrast between the $\mathcal{IV}$ enclosed within the horseshoe region ($\mathcal{IV}_\text{h}$) and the local, approximately Keplerian, $\mathcal{IV}$ ($\approx2\Omega_\text{K}^{-1}$). This process can be repeated several times. In the above equation, $x_\text{h}$ denotes the half-width of the horseshoe region \citep{paardekooperVortexMigrationProtoplanetary2010}, $u_R$ is the local radial velocity of the disc, and quantities with a subscript `p' are evaluated at the planet's location. 

\begin{figure}[t]
   \centering
   \includegraphics[width=\hsize]{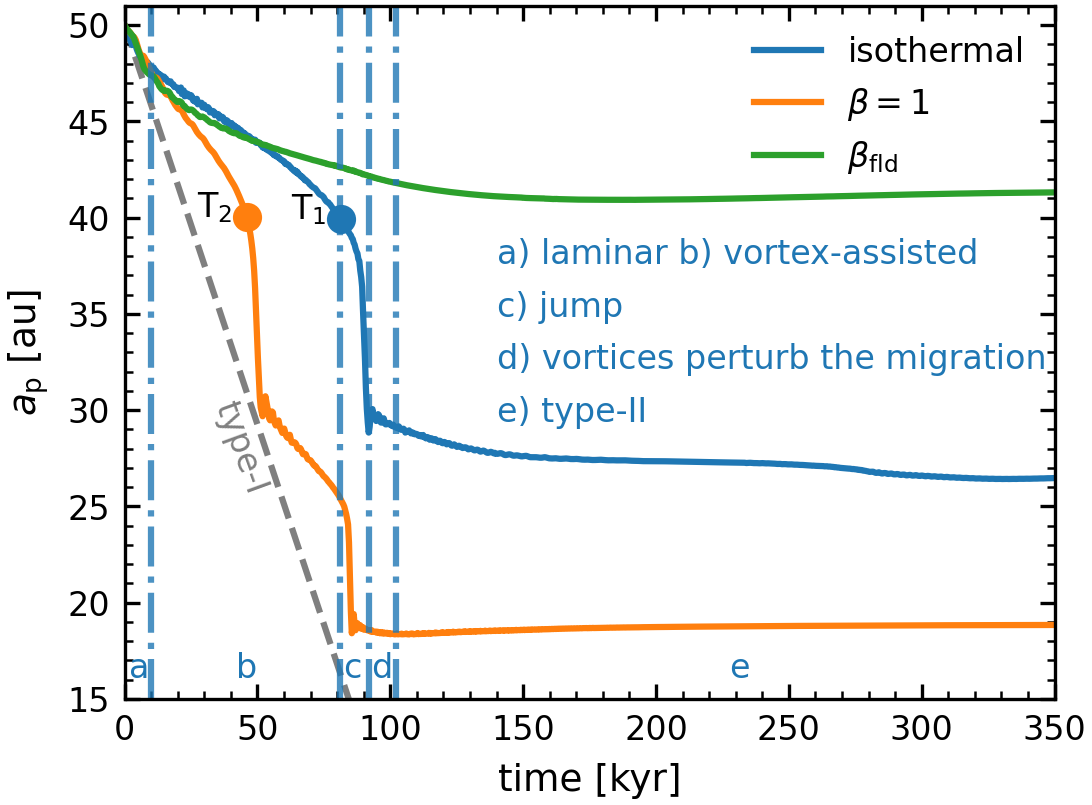}
      \caption{Semi-major axis ($a_{\m{p}}$) evolution of models \texttt{iso}, \texttt{beta}, and \texttt{adapt}. Figure~\ref{vortices iso} shows the inverse vortensity at $\text{T}_{1}$ and $\text{T}_{2}$. The dash-dotted lines indicate the different stages of the migration of model \texttt{iso}. The plot is truncated at 350\,kyr because the planet’s semi-major axis does not change significantly thereafter.}
         \label{semimajor axis}
\end{figure}

Generally, intermittent migration can be further
divided into the following phases, which we illustrate using our fiducial model \texttt{iso} (Fig.~\ref{semimajor axis}). In our models intermittent migration differs slightly from those described in \citetalias{wafflard-fernandezIntermittentPlanetMigration2020}, in particular, as we find that vortex formation is triggered after a jump, more in line with the description of \citetalias{meinersPlanetMigrationALMA}.
\begin{itemize}
\item[a)] First, due to gap
opening by the planet, an $\mathcal{IV}$ maximum forms ahead of the planet ($R < R_{\m{p}}$) and an $\mathcal{IV}$ minimum behind it ($R > R_{\m{p}}$).
\item[b)] The minimum and maximum migrate inwards together with the planet. In this paper, the planet migrates in the vortex-assisted regime.
\item[c)] The planet’s migration is accelerated by small RWI vortices at the outer gap edge. This acceleration enables the planet to catch up with the $\mathcal{IV}$ structures it has generated, which in turn triggers a short episode of runaway migration. All material in the $\mathcal{IV}$ maximum ahead of the planet then enters the horseshoe region, performs U-turns, and accelerates migration (see Eq.~\ref{eq:DCT}), breaking into small vortices during this process. 
\item[d)] This runaway phase ends when all material initially in the $\mathcal{IV}$ maximum ahead of the planet has entered the horseshoe region. After the jump, the planet's semi-major axis fluctuates for a short duration due to the small vortices created during the jump.
\end{itemize}

\subsection{Planetary migration} \label{migration models} 
This section provides an evaluation of the migration behaviour of each simulation listed in Tab.~\ref{tab:sim-list}. We then analyse the post-processed intensity profiles in Sect.~\ref{post processing}.

\subsubsection{Isothermal model} \label{sec:iso}

\begin{figure}[t]
   \centering
   \includegraphics[width=\hsize]{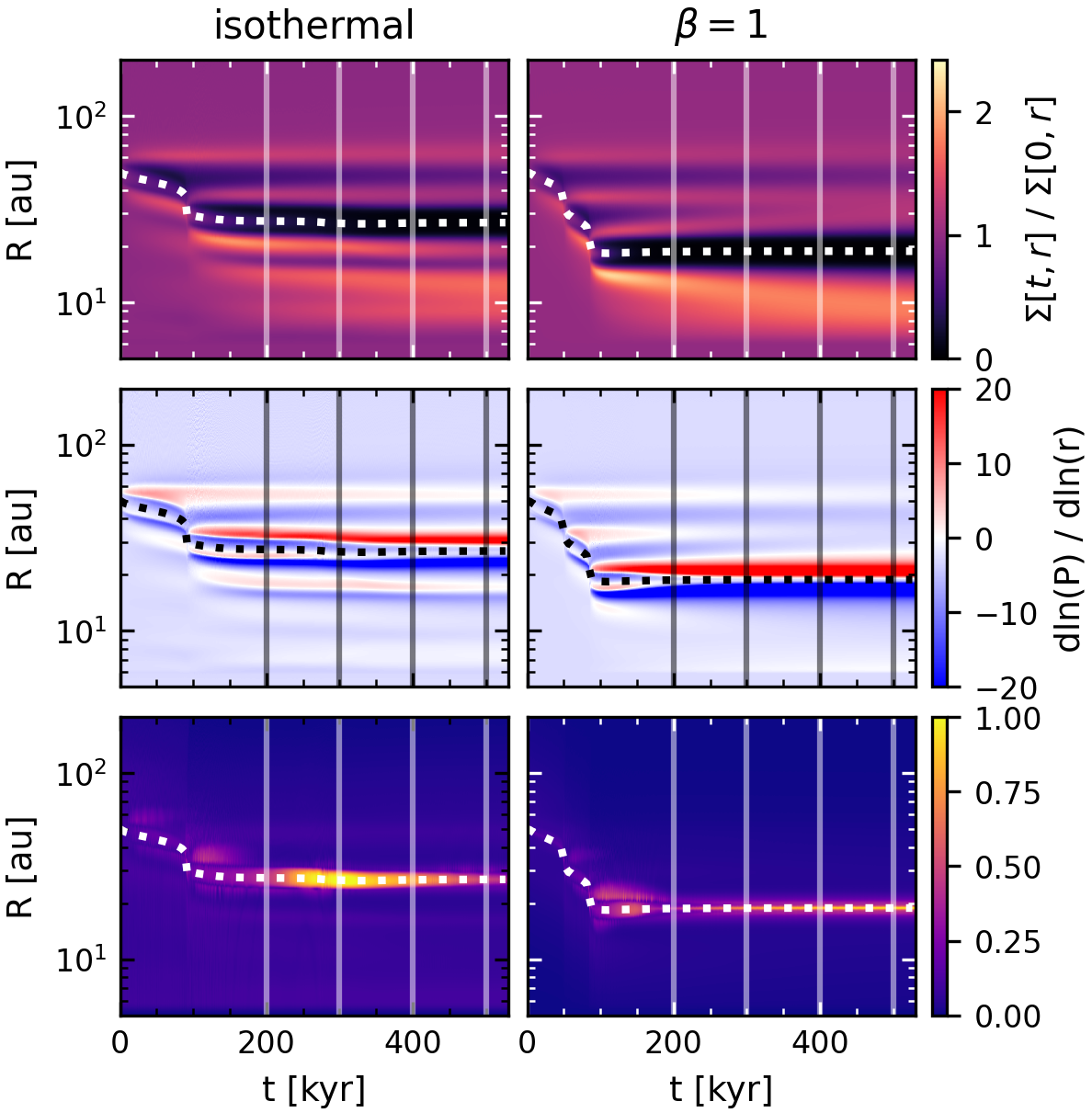}
      \caption{Averaged surface density (top), azimuthally averaged values $dln(P)/dln(r)$ (centre), and standard deviation of $\Sigma$ in the azimuthal direction, normalised to show asymmetries in $\Sigma$ (bottom), for models \texttt{iso} (left) and \texttt{beta} (right) over the span of the whole disc and the duration of the simulation. In the centre panels, white regions at the transition from red to blue (towards larger $R$) indicate local pressure maxima. Transition from blue to red (towards larger $R$) indicate local pressure minima. Light pink regions in the bottom panels indicate large-scale vortices. Values within the horseshoe region cannot be correctly plotted and should be ignored. Dotted lines show the planet's semi-major axis. Vertical lines indicate the times at which we plot the convolved intensity profile in Fig.~\ref{convoluted images}.}
         \label{pressure iso and beta}
\end{figure}

\begin{figure}[t]
   \centering
   \includegraphics[width=\hsize]{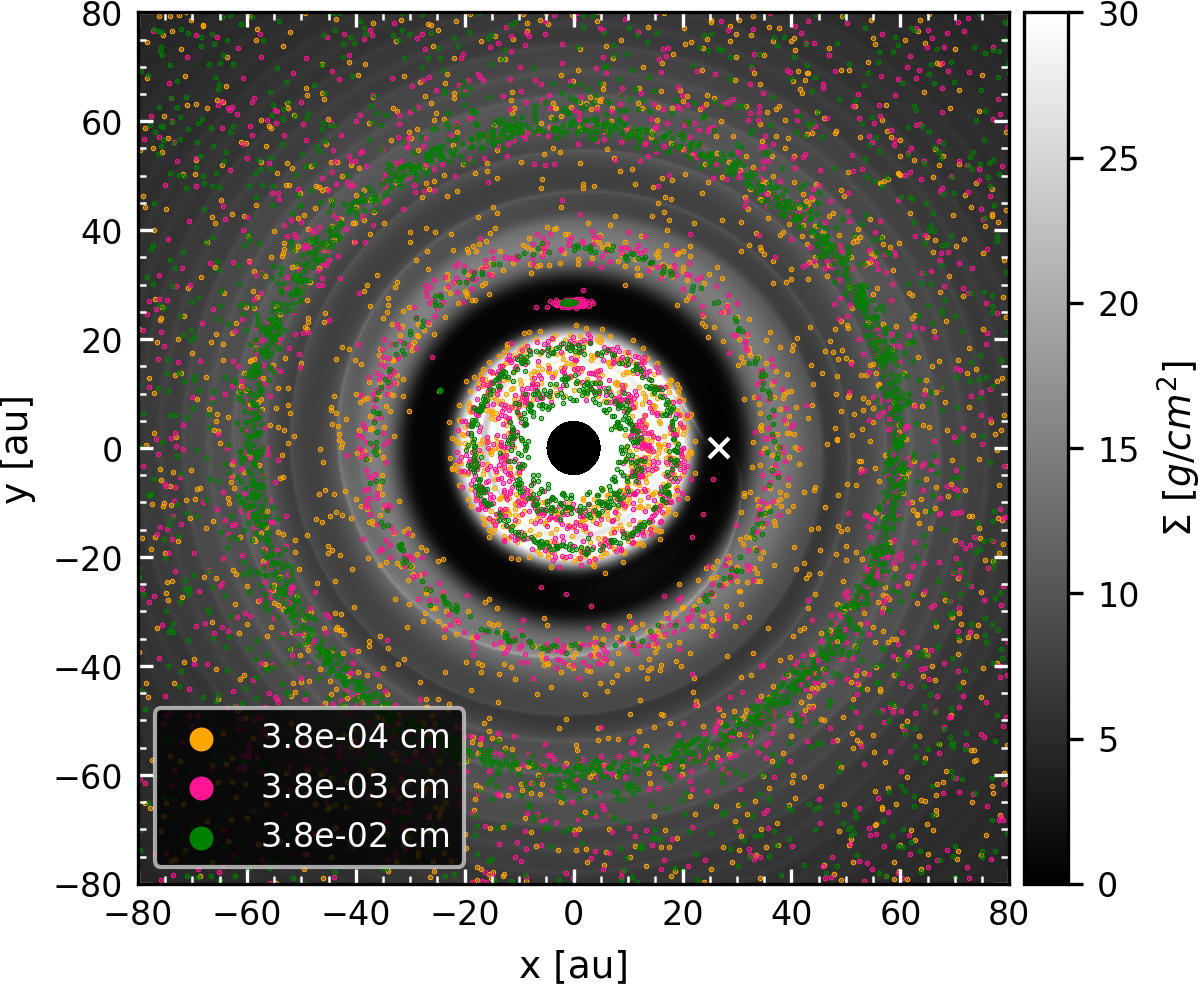}
      \caption{Surface density profile of model \texttt{iso} at $t=500$\,kyr. The inset shows which colours correspond to the respective particle sizes.}
         \label{dust and gas iso}
\end{figure}

\begin{figure}[t]
   \centering
   \includegraphics[width=\hsize]{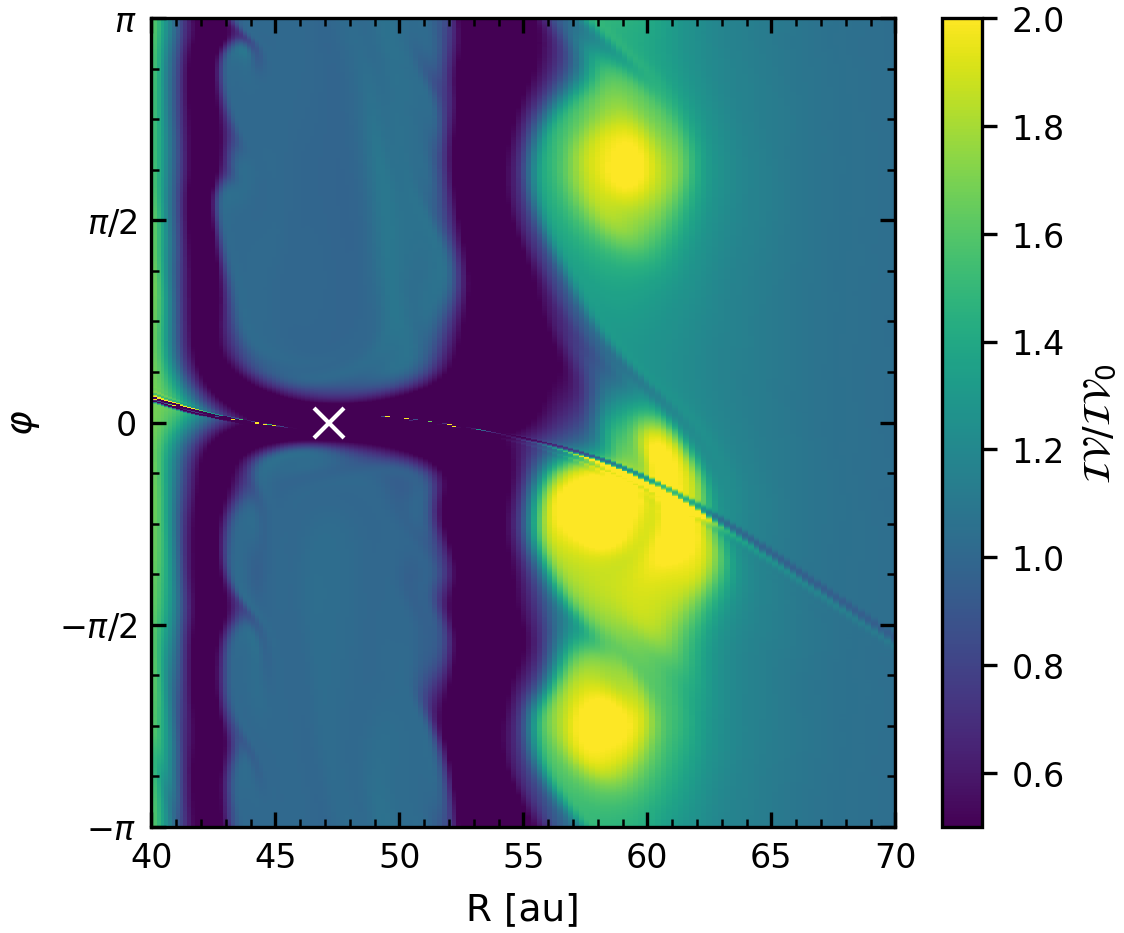}
      \caption{Normalized inverse vortensity of model \texttt{iso} at $t\approx14$\,kyr. The quantity $\mathcal{IV}_{0}$ denotes the initial inverse vortensity of the disc. Three large-scale vortices are visible around $R = 60$\,au. The white cross marks the planet's location.}
         \label{iso vortex}
\end{figure} 

First, we analyse the simulation with the most simplified equation of state (EOS): \texttt{iso}. This simulation confirms the results of \citetalias{meinersPlanetMigrationALMA}, \citetalias{mcnallyMigratingSuperEarthsLowviscosity2019}, and \citetalias{wafflard-fernandezIntermittentPlanetMigration2020} and serves as the fiducial model of this paper.

Fig.~\ref{semimajor axis} shows the evolution of the semi-major axis of the planet. The planet starts its migration in the classical type-I regime for a very short period of time, $\sim 3$\,kyr, during which no significant gap opens. It then transitions into the feedback regime and slows down. Shortly after, the planet enters the vortex-assisted regime, partially accelerating its migration again. Around $t \sim 81$\,kyr, the planet undergoes a migration jump as described in \citetalias{mcnallyMigratingSuperEarthsLowviscosity2019}. After the jump, at $R_{\m{p}} \approx 26$\,au, the planet transitions to the type-II regime, carving a substantially deeper gap as it becomes significantly super-thermal, with $M_{\m{p}} = 3.4 \, M_{\m{th}}$ at this radial location \citepalias{ziamprasHaltingMigrationSuperEarths2025}. Around $t = 320$\,kyr, the planet transitions to slow outward migration.

The migration jump and the planet's ability to open a gap create pressure maxima and minima in the disc (\citetalias{mcnallyMigratingSuperEarthsLowviscosity2019}, \citetalias{wafflard-fernandezIntermittentPlanetMigration2020}), which act as dust traps (\citealt{Pinilla-et-al-2012}). The centre-left panel in Fig.~\ref{pressure iso and beta} shows the azimuthally averaged radial pressure gradient for the model
\texttt{iso}, where $P$ is the  vertically integrated pressure. We observe three distinct pressure maxima ($R_{\m{max}} \approx 60,\,37$, and $20$\,au) and four additional minima ($R_{\m{min}} \approx 48,\,32,\, 25$, and $17$\,au). The planet-induced density variation causes a corresponding change in pressure, shifting the pressure maxima to the locations of the density maxima (\citealt{goldreichExcitationDensityWaves1979, Crida-et-al-2006}). Therefore, the pressure minima coincide with the gaps carved by the planet. Here, the minima at $R_{\m{min}} \approx 32$ and $25$\,au lie within the planet's primary gap. The pressure maxima coincide with the rings (Fig.~\ref{dust and gas iso}). 

\begin{figure}[t]
   \centering
   \includegraphics[width=\hsize]{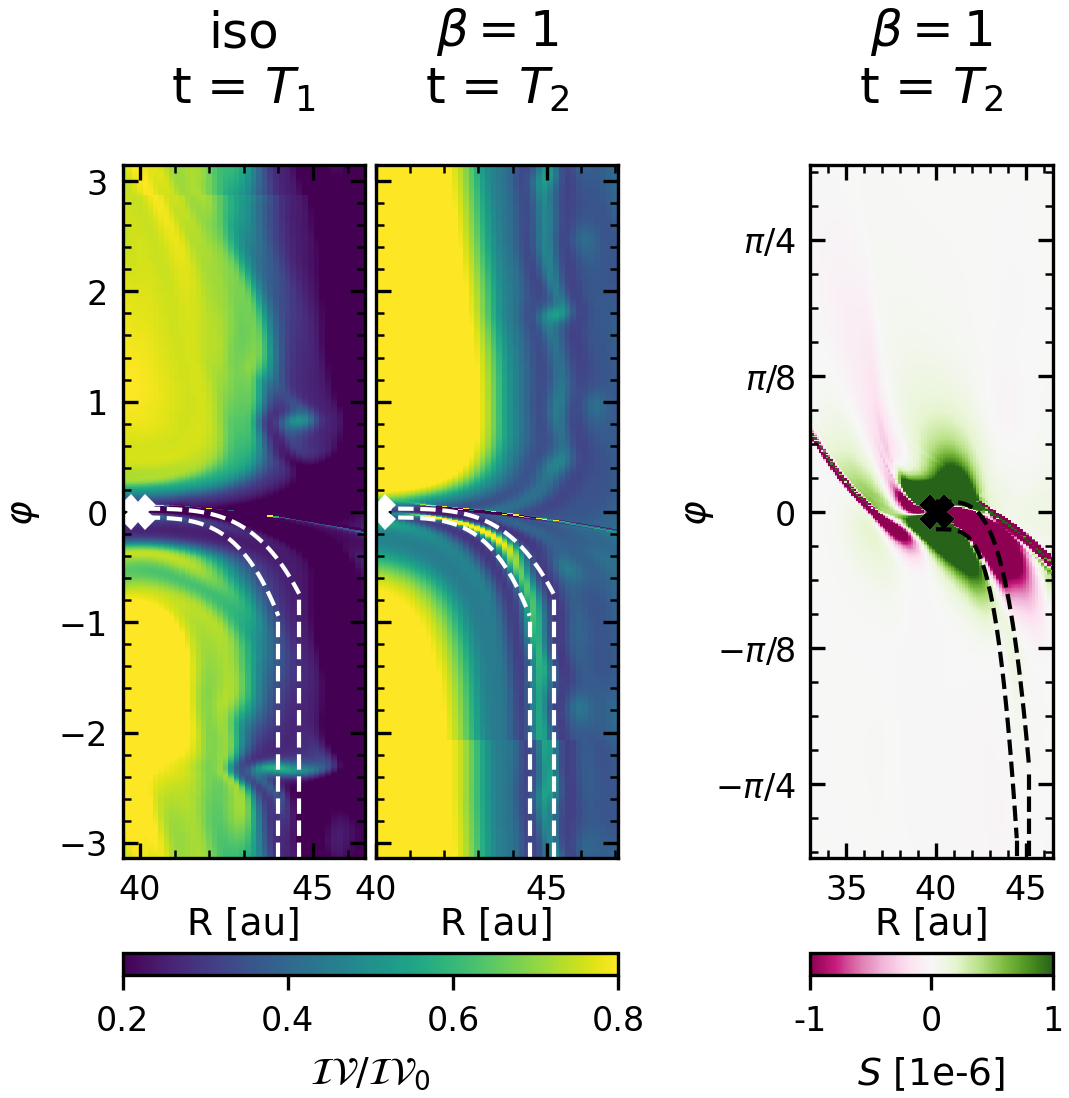}
      \includegraphics[width=\hsize]{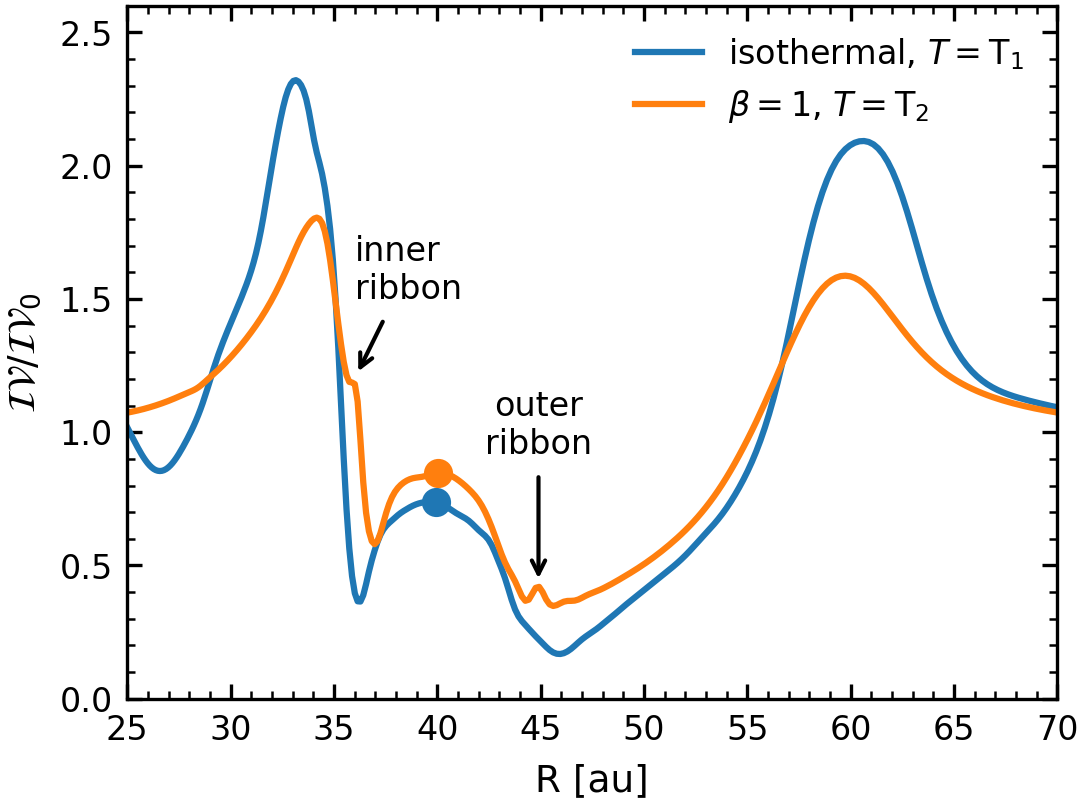}
      \caption{Small-scale vortices at the outer horseshoe edge of model \texttt{iso} (left) and model \texttt{beta} (centre) during migration in the vortex-assisted regimes. Right: Baroclinic forcing (in cgs units) of model \texttt{beta} in the vicinity of the planet. The quantity $\mathcal{IV}_{0}$ denotes the initial inverse vortensity of the disc. Without baroclinic forcing, they appear as maxima in the $\mathcal{IV}$ profile around $R \approx 44.5$ au. With baroclinic forcing, we observe `ribbons' of high and low $\mathcal{IV}$, where high-$\mathcal{IV}$ material breaks into vortices around $R \approx 46.5$ au. These structures are indicated by dotted lines. We also observe an asymmetry in the baroclinic forcing term. The white and black crosses marks the planet's position. Bottom: Averaged $\mathcal{IV}$ profiles at the same points in time. }
         \label{vortices iso}
\end{figure}

\begin{figure}[t]
   \centering
   \includegraphics[width=\hsize]{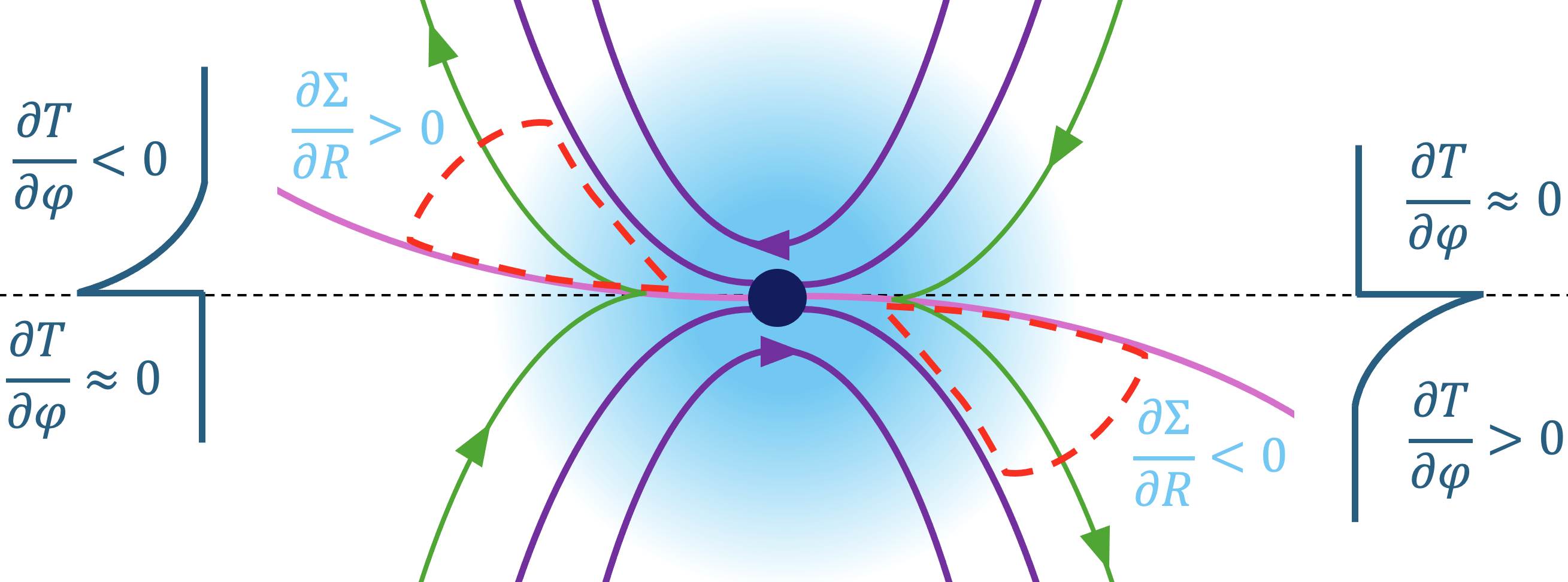}
      \caption{Schematic illustration explaining the baroclinic forcing contribution which creates the `ribbon' of high $\mathcal{IV}$ in model \texttt{beta}, as shown in Fig.~\ref{vortices iso}. Its shows the planet (dark blue dot), the orbiting material in the horseshoe region (purple), the spirals created by the planet (pink), and the unbound material (green). The red-outlined areas highlight the regions of negative baroclinic forcing.}
         \label{explanation BF}
\end{figure}

As seen in \citetalias{meinersPlanetMigrationALMA}, large scale vortices form but have a very short lifetime. Fig.~\ref{iso vortex} illustrates their appearance in the $\mathcal{IV}$ profile. We detect three large-scale ($\sim 8$\,au radial extend) vortices, which appear as local maxima in the $\mathcal{IV}$ profile. Over the next 10\,kyr, these three vortices merge into a single vortex, which dissipates after 100\,kyr (Fig.~\ref{pressure iso and beta}). We observe an additional large-scale vortex created after the migration jump between $40$--$30$\,au, at $R \approx37$\,au, with a lifetime of $\sim$90\,kyr. Both vortices are fully decoupled from the planet. They eventually smear out azimuthally, forming rings of low vortensity and, importantly, preserving the underlying pressure maxima.

The gap at $R \approx 48$\,au is carved by the planet during its slow inward migration up to  $t \sim 81$\,kyr. Even after the migration jump, the initial primary gap and resulting pressure profile dissipate slowly due to the low viscosity and remain visible in the gas surface density profile at $t=500$\,kyr. Figures~\ref{pressure iso and beta}~and~\ref{dust and gas iso} additionally show that the planet carves a secondary gap at $R_{\m{max}} \approx17$\,au. This gap is created by shocks from the secondary and tertiary spirals launched by the planet (\citealt{Bae-Zhu-2018a, Bae-Zhu-2018b}) and has been observed by \citet{zhangDependenceStructurePlanetopened2024} and \citetalias{meinersPlanetMigrationALMA}.

While the gas surface-density profile and dust-particle distribution share the same features, these features appear amplified in the dust (Fig.~\ref{dust and gas iso}). This includes material at the co-orbital Lagrange point L5, which is not visible in the gas surface-density profile. We further observe that the outer ring is predominantly composed of faster-drifting $0.3$\,mm-sized particles that are not located within the gaps, whereas smaller particles follow the gas across pressure maxima (\citealt{Weidenschilling-1977, NAKAGAWA1986375, zhangDiskSubstructuresHigh2018}).

\subsubsection{\texorpdfstring{Constant $\beta$}{Constant beta}}\label{sec:const beta}

We now introduce cooling with a constant cooling timescale $\beta=1$. Fig.~\ref{semimajor axis} shows the semi-major axis evolution of model \texttt{beta}. The planet starts carving a gap, producing the same $\mathcal{IV}$ minima and maxima structures as in model \texttt{iso}. In addition to the Rossby-wave unstable partial gap, cooling drives significant baroclinic forcing at the gap edges \citepalias{ziamprasHaltingMigrationSuperEarths2025}. The asymmetric baroclinic forcing profile produces three ribbons of high and low $\mathcal{IV}$ at the outer horseshoe edge, shown in the right panel of Fig.~\ref{vortices iso}, as opposed to model \texttt{iso} (left panel of the same figure). Further analysis (see also Appendix~\ref{appendix B}) shows that these ribbons are created by gas outside the horseshoe orbit interacting with the planetary shocks. Figure~\ref{explanation BF} shows a schematic sketch of the contribution term $\partial \Sigma/\partial R \cdot \partial T/\partial \phi$ of the baroclinic forcing term $\mathcal{S}$ (Eq.~\eqref{baroclinic forcing}) in the red-outlined area. The planet accumulates mass around its Hill sphere, which determines $\partial \Sigma/\partial R$. Meanwhile, unbound gas outside the horseshoe orbits is heated by the planetary shock and subsequently cools, determining $\partial T/\partial \phi$. Combining both terms results in a negative baroclinic forcing contribution, which produces the low vortensity ribbon in model \texttt{beta}. For an isothermal EOS, $\partial T/\partial \phi = 0$, setting this baroclinic forcing contribution to 0. The baroclinic forcing acts as a very efficient vortensity `sink', to the point that the gradient relative to the background becomes large enough for the flow to quickly break into vortices. These vortices refill the outer gap faster than those seen in the \texttt{iso} model. Additionally, the term $\partial \Sigma/\partial \phi \cdot \partial T/\partial R $ in the baroclinic forcing drives vortensity inside the horseshoe region and thus contributes to inward migration, as in \cite{ziamprasMigrationLowmassPlanets2024}. This is further discussed in Appendix~\ref{appendix B}.

\begin{figure}[t]
   \centering

   \includegraphics[width=\hsize]{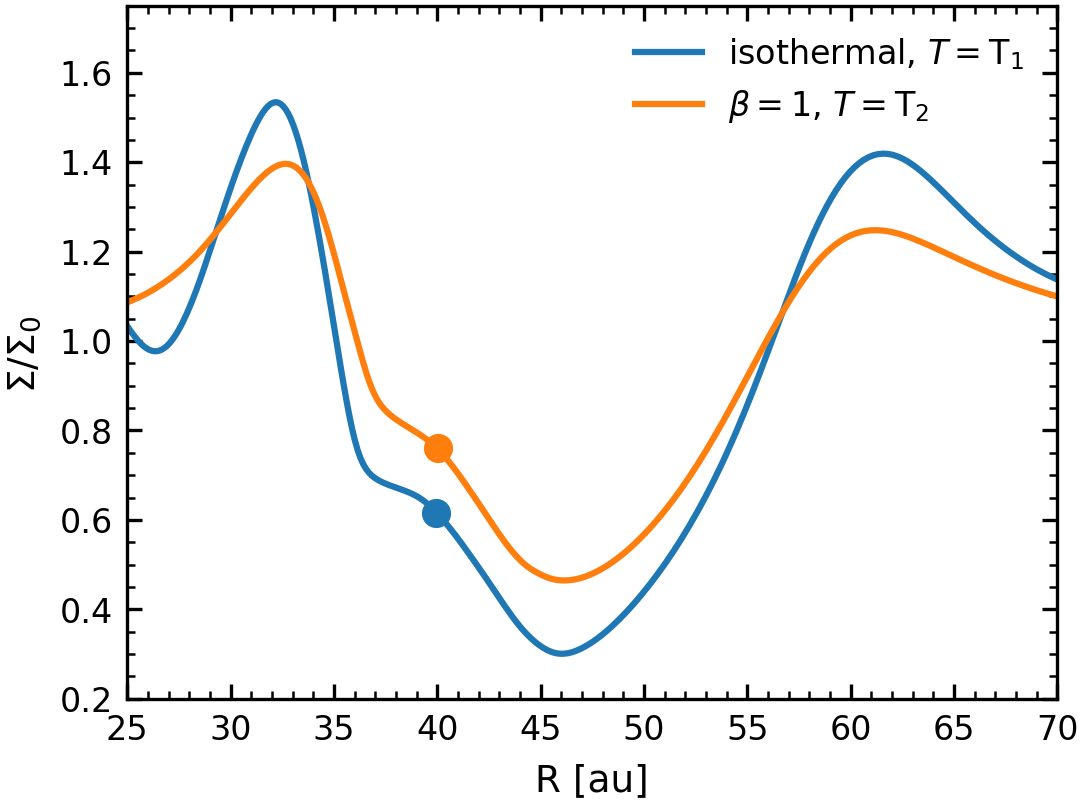}
      \caption{Normalized, azimuthally averaged gas surface density profile of models \texttt{iso} and \texttt{beta} at $t = \text{T}_{1}$ and $\text{T}_{2}$. The dot indicates the planet's position.}
         \label{surf_dens_compare}
\end{figure}

The planet migrates in the vortex-assisted regime and undergoes its first intermittent migration jump at $t \sim 46$\,kyr. At this time, the planet is located at $R_{\m{p}} = 40$\,au, the same radius at which model \texttt{iso} enters the intermittent migration. In both simulations, the planet ends the migration jump at $R_{\m{p}} \approx 30$\,au. Whether and when the planet undergoes a jump is determined by its ability to catch up with its own gap structure. The structure of the gaps depends on the shock distance, which is proportional to $c_{s}$ (\citealt{goodmanPlanetaryTorquesViscosity2001}). \cite{mirandaPlanetDiskInteraction2020} determined an effective $c_{s}$, which is very close to the isothermal sound speed for $\beta =1$. Therefore, we expect that both planets open a similar gap, which is confirmed by Fig.~\ref{surf_dens_compare}. Figures 4 and 5 in \cite{ziampras-etal-2026} also show that the angular-momentum deposited in the gap is similar for isothermal and $\beta = 1$ simulations. Model \texttt{beta} presents a shallower gap, because it was plotted at an earlier time ($\text{T}_{1} > \text{T}_{2}$).

The planet re-enters the vortex-assisted regime after the first jump and carves a deeper gap than before the jump. The process repeats: the planet modifies its surroundings to create a new $\mathcal{IV}$ minimum and maximum around it, and the RWI vortices remain strong enough to significantly refill the now deeper primary gap, maintaining the inward migration. As a result, the planet undergoes a second intermittent migration jump at $t = 82$\,kyr, covering another 6.6\,au in 4\,kyr.

After the second jump, the planet reaches the type-II regime and migrates outward slowly from $R_{\m{p}} \approx 18$\,au. Here, $M_{\m{p}} = 4.49\,M_{\m{th}}$, and the planet mass is super-thermal, as expected (\citealt{ziamprasHaltingMigrationSuperEarths2025}). The planet's primary gap depth decreases rapidly, which is expected, since the primary gap opening is most efficient for $\beta = 1$ (\citet{mirandaPlanetDiskInteraction2020}).

\begin{figure}[t]
   \centering
   \includegraphics[width=\hsize]{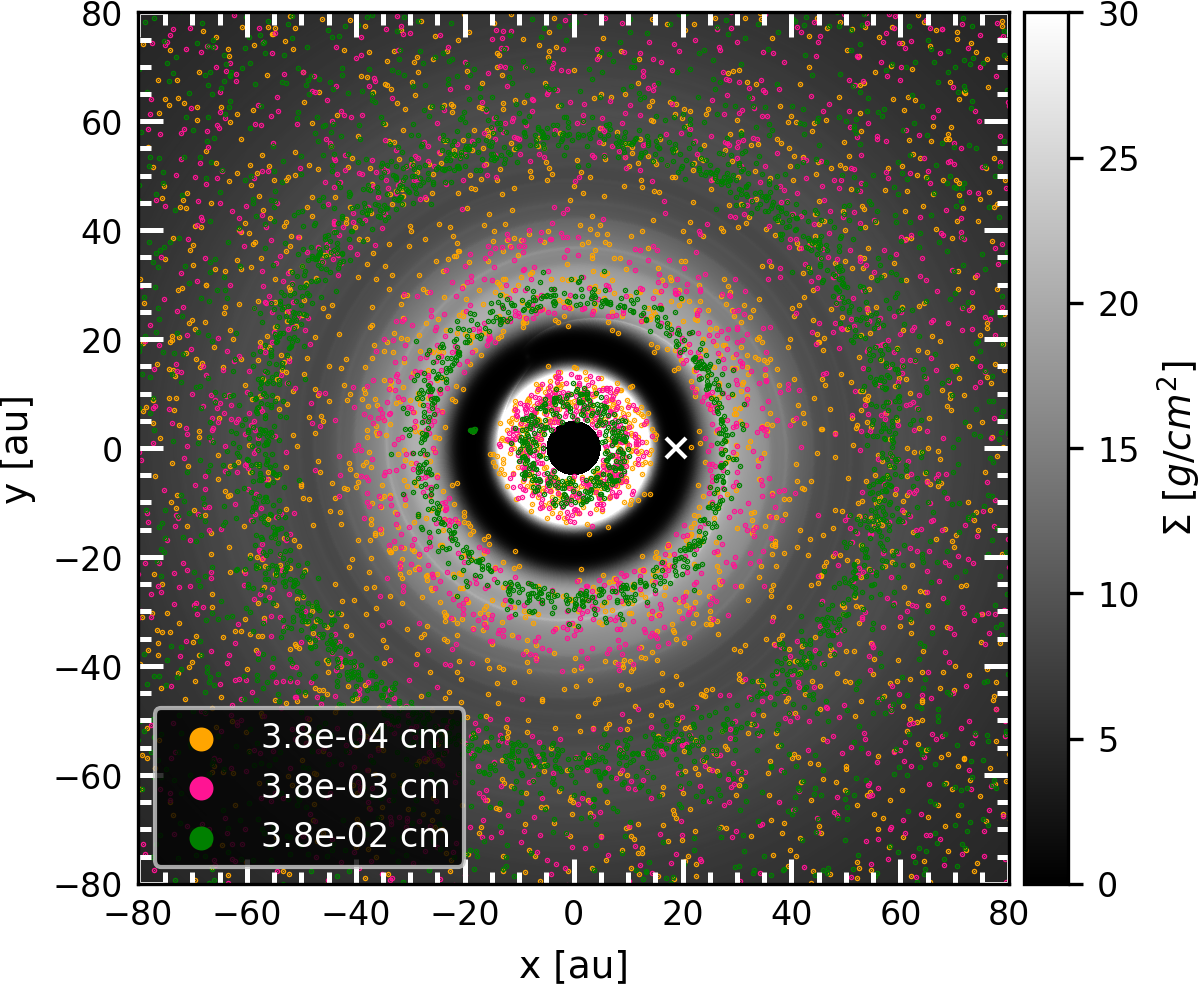}
      \caption{Surface density profile with dust for model \texttt{beta} at $t=500$\,kyr. The inset shows which colours correspond to the respective particle sizes.}
         \label{dust and gas beta}
\end{figure}

We observe that the pressure maxima created by the planet at $R_{\m{max}} = 37$\,au and $R_{\m{max}} = 24$\,au begin to overlap around $t\sim 300$\,kyr, since the gap edge at $R = 24$\,au shifts towards higher radii (Fig.~\ref{pressure iso and beta}). Therefore, the gas surface density and particle distribution in Fig.~\ref{dust and gas beta} show only the outermost and innermost rings ($R_{\m{max}} \approx 60$ and $28$\,au) and gaps ($R_{\m{min}} \approx 47$\,au and $18$\,au), even though the planet undergoes two intermittent migration phases. Similar to model \texttt{iso}, the dust distribution shows similar but more pronounced features compared to the gas surface density, with larger grains populating the rings. No secondary gap is opened interior to the planet's orbit, as expected for $\beta = 1$ (Fig.~\ref{pressure iso and beta}).

During the first jump, when the planet moves from 40\,au down to 30\,au, a large vortex forms around $R\approx 35$\,au, with a lifetime of $\sim 50$\,kyr, which is not coupled to the planet. The second jump (the planet moves from 25\,au to 18\,au) generates two large-scale vortices at $R \approx 24$\,au and $R \approx 16$\,au. All three exhibit the same $\mathcal{IV}$ structure as the vortex in Fig.~\ref{iso vortex} . While the vortex at $R \approx 16$\,au only has a lifetime of $\sim 30$\,kyr, the vortex at $R \approx 24$\,au survives significantly longer, for 90\,kyr. The latter may be periodically perturbed by the planet, which delays its dissipation.

\subsubsection{Adaptive $\beta$}

\begin{figure}[t]
   \centering
   \includegraphics[width=\hsize]{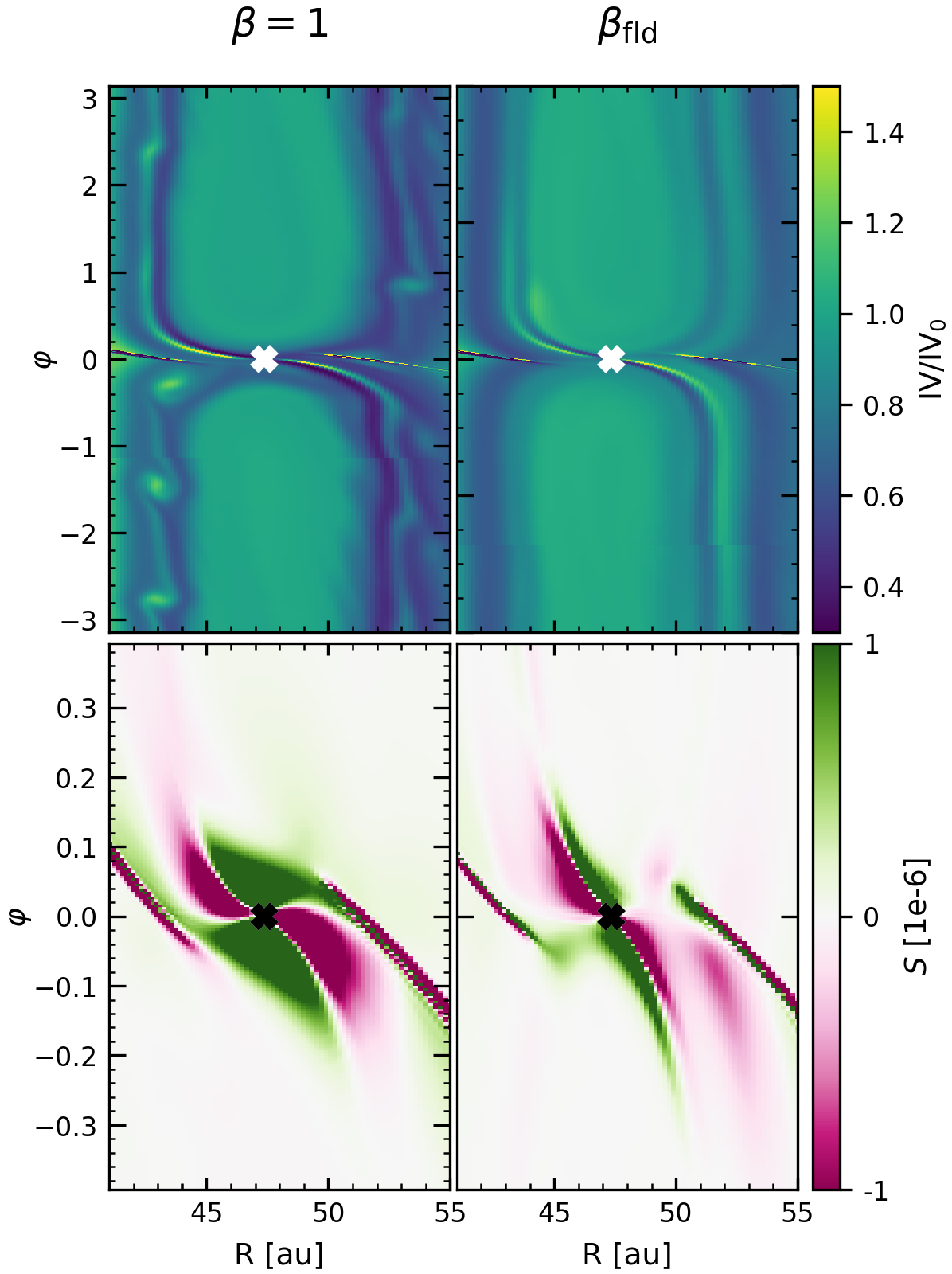}
      \caption{Normalized inverse vortensity profile and baroclinic forcing $S$ (in cgs units) in the horseshoe region of models \texttt{beta} (left) and \texttt{adapt} (right) at $t \approx 10$\,kyr. Although both simulations include cooling, model \texttt{beta} exhibits a significantly more asymmetric baroclinic forcing term , which produces the characteristic ribbons of high and low $\mathcal{IV}$, which break into vortices. The more symmetric baroclinic forcing in \texttt{adapt} does not produce this $\mathcal{IV}$ structure. The white cross marks the planet's position.}
         \label{beta_adapt_compare}
\end{figure}

Next, we introduce an adaptive cooling timescale $\beta_{\m{fld}}$, the model closest to the full radiative disc description. Fig.~\ref{semimajor axis} shows the semi-major axis evolution of model \texttt{adapt}. After opening a sufficiently deep gap, the planet initially migrates in a type-I like regime, with a starting value of $\beta_{\m{fld}} \approx 0.15$. The disc structure changes, creating an $\mathcal{IV}$ maximum ahead of the planet and a minimum behind it.

After 10\,kyr, this changes, and the planet enters the vortex-assisted regime with an asymmetric baroclinic forcing term. Baroclinic forcing is much weaker than in the $\beta=1$ case, with only small, sometimes barely visible vortices in the $\mathcal{IV}$ profile (Fig.~\ref{beta_adapt_compare}). Additionally, the baroclinic forcing term becomes asymmetric only once the planet has already opened a significant gap. Both effects together result in slow migration, and the planet does not catch up with the created $\mathcal{IV}$ structures. Hence, the planet does not undergo a migration jump. We observe an additional negative baroclinic forcing feature at $R \approx 52$\,au in model \texttt{adapt}, likely caused by the stronger temperature dependence of $\beta_{\text{fld}}$.

At $t \sim 180\,\text{kyr}, R_{p} \sim 41$\,au, the planet enters the type-II regime but migrates outward. At this point, $M_{\m{p}} = 2.4 M_{\m{th}}$ and $\beta_{\m{fld}}$ has an approximate value of $\beta_{\m{fld}} \approx 0.146$, which is close to the starting value. We observe no large-scale vortices over the duration of the simulation.

Although $\beta_{\m{fld}}$ varies across the disc, the planet halts its migration in the transition between the optically thin and optically thick disc regimes, where the dependence on $\Sigma$ is weak (Fig.~\ref{beta fld profile}) and does not trigger a runaway process. Even though the planet stalls its migration, baroclinic forcing continues to increase in strength due to the descending gap depth and the increasing steepness of the gap edge. At $t \sim 450$\,kyr the small-scale vortices produced by baroclinic forcing become strong enough to appear in the $\mathcal{IV}$ profile (Fig.~\ref{vortices iso}, centre plot) and the planet could soon re-enter the vortex-assisted regime. This is supported by the migration track, which now no longer follows a smooth curve, but fluctuates. 

Nevertheless, the planet did not undergo intermittent migration in this simulation; consequently, only the primary and secondary gaps carved by the planet are present. Since this work focuses on planets that experience migration jumps and on their resulting structures, we reduced the planet mass from $M_{\m{p}} = 100\, \text{M}_{\oplus}$ to $M_{\m{p}} = 62\, \text{M}_{\oplus}$, corresponding to $M_{\m{p}} \approx 1.3 M_{\m{th}}$. We name this new simulation \texttt{adapt-low-mass}. While we expect the migration behaviour to change, we note that $\beta_{\m{fld}}$ does not depend on the planet's mass, enabling a fair comparison between the two models.

\subsubsection{Adaptive $\beta$ with a lower-mass planet} \label{migration 67 me models}

\begin{figure}[t]
   \centering
   \includegraphics[width=\hsize]{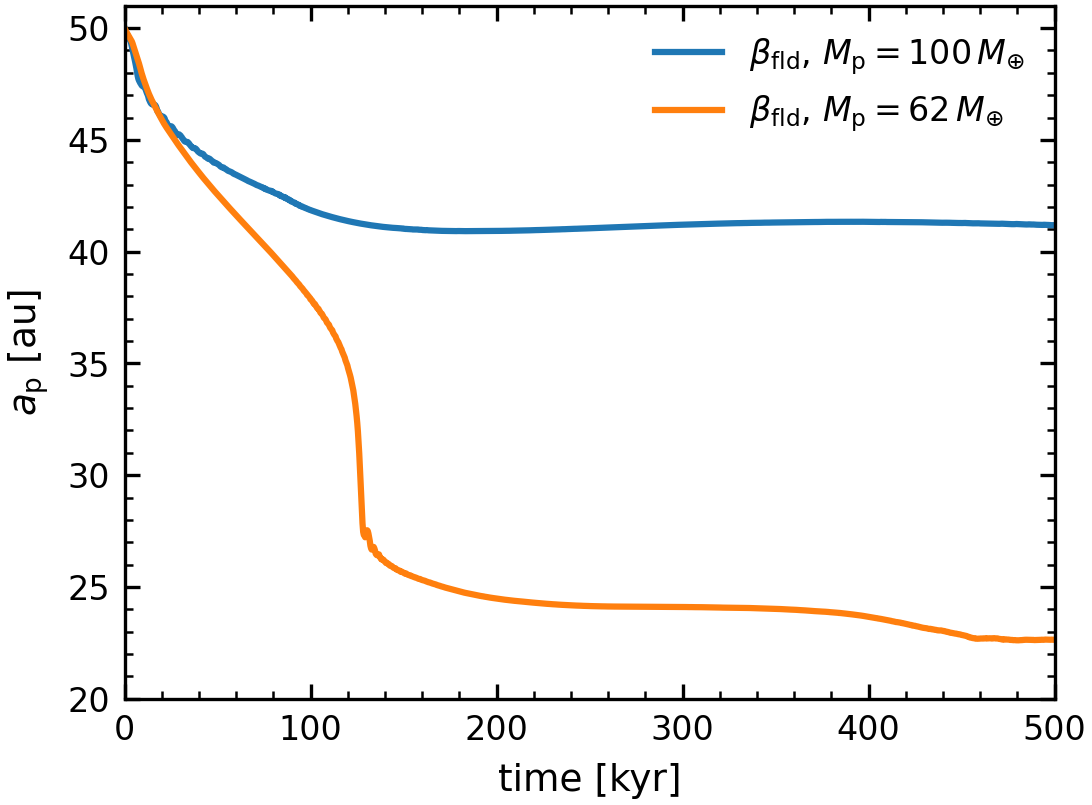}
      \caption{Semi-major axis ($a_{\m{p}}$) evolution of models \texttt{adapt-low-mass} and \texttt{adapt}.}
         \label{semimajor axis fld 62}
\end{figure}
\begin{figure}[t]
   \centering
   \includegraphics[width=\hsize]{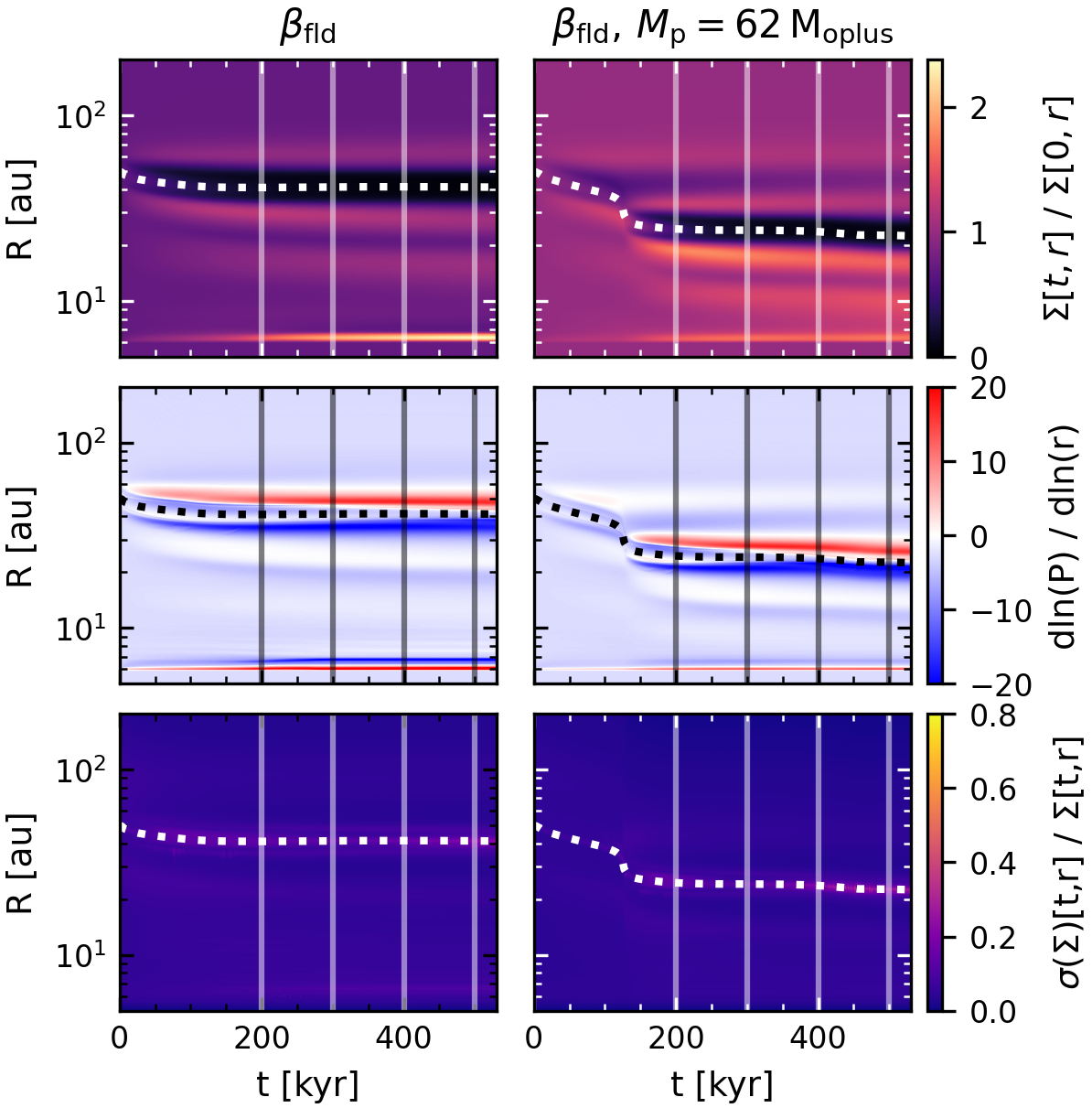}
      \caption{Averaged surface density (top), azimuthally averaged values $dln(P)/dln(r)$ (centre), and asymmetries in $\Sigma$ (bottom) of models \texttt{adapt} (left) and \texttt{adapt-low-mass} (right) over the span of the whole disc and the duration of the simulation. Due to the variable cooling timescale, a temperature spike forms in front of the inner damping zone. This does not influence the planet's migration.}
         \label{pressure fld}
\end{figure}

Fig.~\ref{semimajor axis fld 62} shows the semi-major axis evolution of model \texttt{adapt-low-mass}. The simulation undergoes an intermittent migration jump. As in model \texttt{adapt}, \texttt{adapt-low-mass} starts its migration in the laminar regime and switches to vortex-assisted migration after 15\,kyr. Initially, small vortices at the outer gap edge are only visible when plotting the deviation from the azimuthally averaged $\mathcal{IV}$ profile, and later they also appear in the $\mathcal{IV}$ profile, as seen in Fig.~\ref{vortices iso} (left panel). At later times, we observe the ribbons of high and low $\mathcal{IV}$ ( Fig.~\ref{vortices iso}, centre panel), characteristic of strong baroclinic forcing. We conclude that the planet is in the vortex-assisted regime, primarily driven by baroclinic forcing. These vortices accelerate the planet sufficiently  to trigger a migration jump around $t = 120$\,kyr, covering $7.5$\,au in $9$\,kyr. The planet reaches an equilibrium at $R_{\m{p}} \approx 24$\,au and migrates in the type-II regime, with $M_{\m{p}} = 2.24 M_{\m{th}}$. There, the cooling timescale assumes a value of $\beta_{\m{fld}} = 0.28$. 

Unlike \texttt{adapt}, the planet migrated far enough into the optically thick regime of the disc, where $\beta_{\m{fld}}$ depends strongly on $\Sigma$ (Fig.~\ref{beta fld profile}). Hence, the planet enters a runaway process as described in Sect.~\ref{sec:migration regimes}, reducing its surface density by a factor of 100 over the next 300\,kyr. Meanwhile, small vortices form at the outer gap edge (as shown in Fig.~\ref{vortices iso}, left panel), likely driven by a combination of the sharp radial vortensity contrast from gap opening and baroclinic forcing, as in model \texttt{adapt}. However, since the gap is continuously cleared, the planet remains in the type-II regime until the end of the simulation.

\begin{figure}[t]
   \centering
   \includegraphics[width=\hsize]{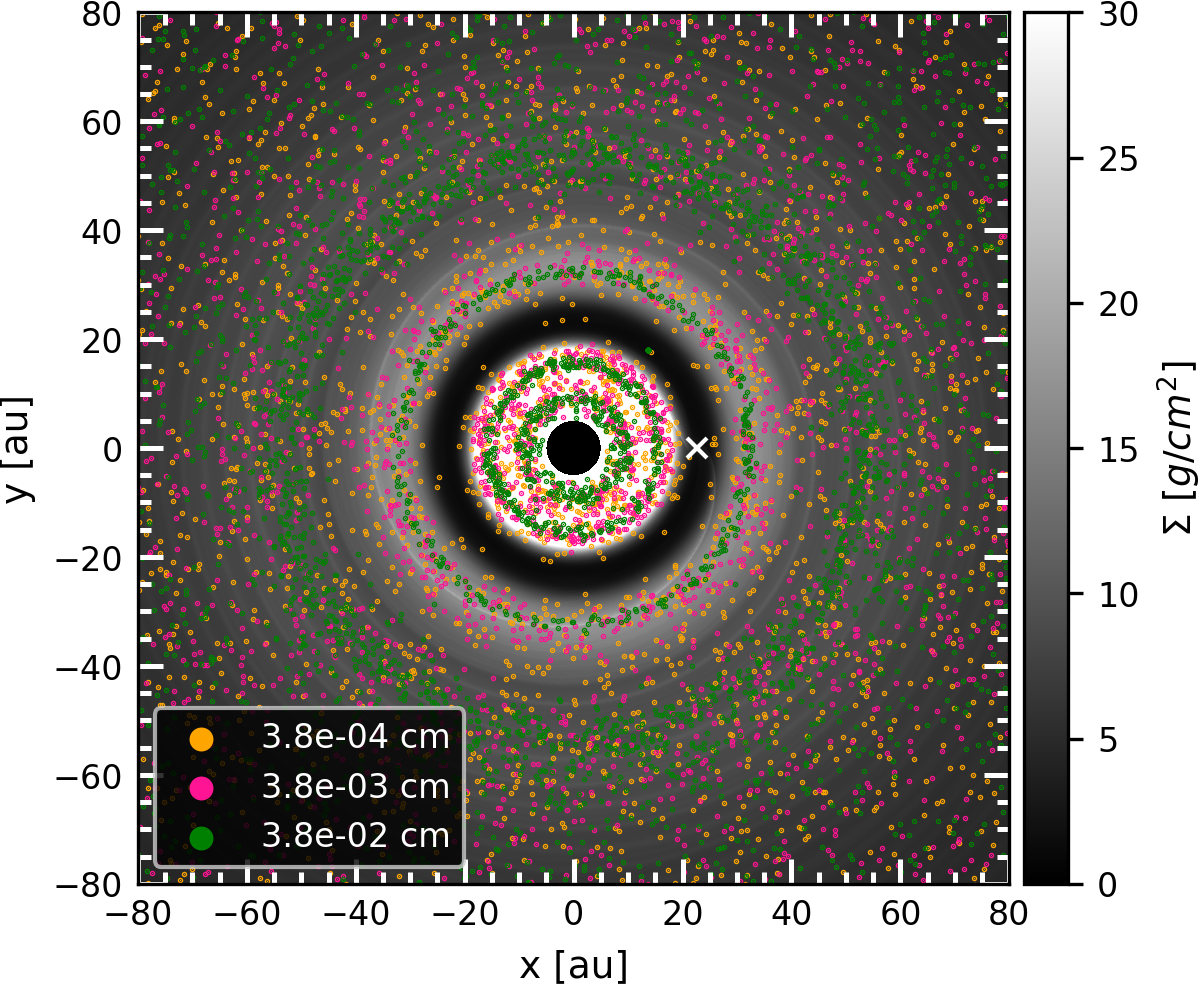}
      \caption{Surface density profile with dust for model \texttt{adapt-low-mass} at $t=500$\,kyr. The inset shows which colours correspond to the respective particle sizes.} 
         \label{dust and gas fld}
\end{figure}

The planet carves three gaps ($ R_{\m{min}} \approx 45 ,\, 24$, and $13$\,au) and three rings ($R_{\m{max}} \approx 60,\, 33 $, and $18$\,au) consistent with the pressure profile. The pressure maxima and minima associated with the innermost gap and ring, caused by the planet's secondary gap, are significantly less pronounced (Fig.~\ref{pressure fld}). Nevertheless, they remain sharp enough to produce visible gaps in the averaged surface density profile (Fig.~\ref{pressure fld}) as well as in the dust particle distribution (Fig.~\ref{dust and gas fld}. The gas and dust distributions are consistent with expectations (Fig.~\ref{dust and gas fld}).

\subsection{Radiative transfer images}\label{post processing}

\begin{figure*}[t]
   \centering
   \includegraphics[width=\textwidth]{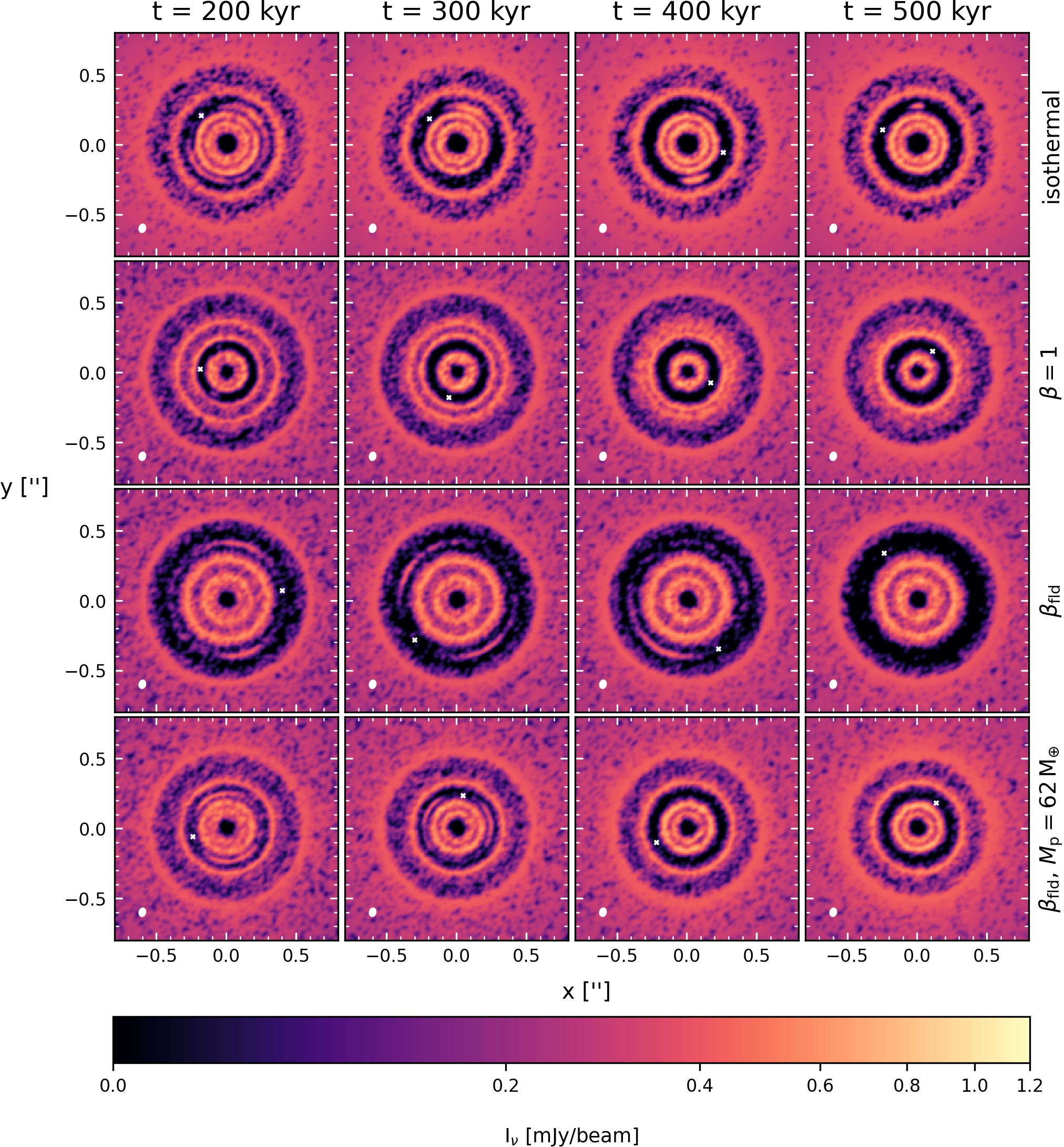}

   \caption{Intensity $I_{\nu}$ of model \texttt{iso}, \texttt{beta}, \texttt{adapt} and \texttt{adapt-low-mass}, convolved with a Gaussian beam at $t=200, 300, 400,$ and $500$\,kyr. The white ellipse represents the beam, and the white cross indicates the planet's position. }
    \label{convoluted images}
\end{figure*}

To produce images similar to those obtained in the DSHARP survey, we modelled the evolution of dust grains alongside the gas, as described in Sect.~\ref{sec:numerical setup}. We then binned the super-particles onto a surface density grid using the \texttt{python} module \texttt{coord2sigma} \citep{rometschCoords2sigma2023}. We calculated the observed intensity $I_{\nu}$ as
\begin{equation}
       I_{\nu} = (1 - e^{-\tau_\nu})\, B_{\nu}.
\end{equation}
Here, $B_\nu(T)$ denotes the black-body intensity profile evaluated at the gas temperature $T_\mathrm{gas}$. We used the DIANA standard opacities for each dust species, computed with \texttt{optool} \citep{dominikOpToolCommandlineDriven2021} at $\lambda \approx 1.25\,\m{mm}$, the wavelength used by the DSHARP survey and ALMA Band 6. We then interpolated the intensity heat map $I_\nu(R,\phi)$ onto a Cartesian grid and convolved with a Gaussian kernel. We assumed a distance of 100 pc and an inclination of 0° for the discs and adopted the configuration used for Elias 20 in DSHARP, with a beam of angular size $\theta_{\m{b}} = 32 \times 23\,\m{mas}$ and a position angle of $\m{PA}_{\m{b}} = 76^{\circ}$ (\citealt{andrewsDiskSubstructuresHigh2018}). This produces observation-like images for model \texttt{iso}, \texttt{beta}, \texttt{adapt}, and \texttt{adapt-low-mass} (Fig.~\ref{convoluted images}). 

Model \texttt{iso} displays several ring and gap features over the duration of the simulation. We also observe material at the co-orbital Lagrange points, which is cleared over time \citep[as also seen in][]{rodenkirchModelingNonaxisymmetricStructure2021}. A small remainder of the dust at the L5 Lagrange point remains visible at 500\,kyr. The planet's continuous gap opening and dust drift causes the rings at $R \approx 37$ and $20$\,au to become brighter as the simulation evolves from $t = 200$\,kyr to $t = 500$\,kyr. Comparing the convolved intensity profile with the dust distribution in Fig.~\ref{dust and gas iso}, the gaps are deeper and more pronounced in the dust picture. Nevertheless, all structures are present in both distributions. 

Compared to model \texttt{iso}, the planet in model \texttt{beta} undergoes two migration jumps instead of one. Fig.~\ref{pressure iso and beta} shows that some pressure maxima merge over the duration of the simulation due to the shift of the gap edge at $R \approx 24$\,au. As shown in Fig.~\ref{dust and gas beta}, this influences the dust particle distribution. Figure~\ref{convoluted images} also shows the merging of the inner ring and the gap structure, leaving only two distinct gaps at $t = 500$\,kyr. We additionally observe that the gap at the planet's final location is almost cleared at $t=200$\,kyr, with very little material remaining at the co-orbital Lagrange points. The planet does not open a secondary gap, as expected for $\beta = 1$ \citep{mirandaPlanetDiskInteraction2020, ziamprasModellingPlanetinducedGaps2023,zhangDependenceStructurePlanetopened2024}. 

The planet in model \texttt{adapt} does not undergo a migration jump; hence, the convolved intensity profiles show only the primary and secondary gap around the planet and interior to its orbit, respectively. Consistent with the other simulations, a substantial accumulation of material is present at the co-orbital Lagrange points, which is progressively cleared by $t=500$\,kyr.

The intensity profile of \texttt{adapt-low-mass} shows three rings and gaps: the two outermost ones are created by the migration jump, while the innermost structures are produced by the planet's secondary gap. The outer two rings remain visible in the intensity profile from their formation until the end of the simulation, 400\,kyr later. The innermost ring and gap only become distinguishable at $t = 300$\,kyr, over 100\,kyr after the planet enters the stalling regime. Although the surface density of the primary gap is drastically reduced \citep[in part due to cooling, e.g.][]{mirandaPlanetDiskInteraction2020}, this does not apply to the planet's secondary gap. Unlike in models \texttt{iso} and \texttt{beta}, there are no large vortices, but model \texttt{adapt-low-mass} still creates pressure maxima. Additionally, the gap at the planet's final location ($R_{\m{min}} \approx 24$\,au) still contains some dust at the co-orbital Lagrange points up to $t = 350$\,kyr, although its surface density decreases rapidly over time. This is due to dust drift. Comparing the particle distribution in Fig.~\ref{dust and gas fld} with the intensity profile in Fig.~\ref{convoluted images}, the planet's secondary gap is less visible, while asymmetries in the ring at $R_{\m{max}} = 18$\,au become more apparent.

This analysis shows that all three simulations that undergo a migration jump produce observable dust structures, which remain visible over the full duration of the simulations. This suggests that we may be able to explain the dynamical history of some planetary systems that show similar structures to those in our simulations.

We note that the `noise' in the synthetic observations arises from the binned particle distribution in the simulation output and is unrelated to instrument noise (which would be negligible on this scale). For the same reason, we caution that apparent asymmetries within rings should not be interpreted as actual non-axisymmetric features.\\

We find that planets in simulations with an isothermal EOS, constant $\beta$-cooling, or $\beta_{\m{fld}}$ can experience one or several migration jumps. These jumps create pressure maxima and minima in the disc, which act as dust traps that translate into visible rings and gaps in the intensity profile. Most of them remain visible at $t =500$\,kyr. We changed the planet mass in model \texttt{adapt}, because the planet with $M_{\m{p}} = 100\, \text{M}_{\oplus}$ did not undergo an intermittent migration jump.\\

\section{Discussion} \label{sec:discussion}

In this section we evaluate our results and their implications. We discuss the limitations of our method and possible ways to improve it. 

\subsection{Planet migration}\label{discussion:migration}

We observe that regardless of the chosen EOS, the planet migrates in the vortex-assisted regime and may undergo one or several migration jumps. At its initial radial location, the planet has a mass of $M_{\m{p}} \sim 2\,M_{\m{th}}$ in models \texttt{iso}, \texttt{beta}, and \texttt{adapt} and $M_{\m{p}} \sim 1.3\,M_{\m{th}}$ in model \texttt{adapt-low-mass}. Since model \texttt{adapt} did not experience a migration jump, while models \texttt{beta} and \texttt{adapt-low-mass} did, we infer that the transition between the intermittent and type-II migration regimes depends on both planet mass and the local cooling timescale. This implies that modelling observed substructures in the context of migrating planets should be carried out carefully, including a prescription for cooling.

We also carried out models with $\beta$ values of $0.01, 3$, and $7.5$, whose migration tracks are shown in Appendix~\ref{appendix A}. The simulations with $\beta = 0.01$ and $3$ also undergo intermittent migration when they reach the same radial location as \texttt{iso} and \texttt{beta}. This is most likely related to these models having comparable gap structures, due to the similar angular momentum flux profiles for $\beta=1$ and $3$ (\citealt{ziampras-etal-2026}). The simulation with $\beta = 7.5$ does not experience any jumps and has a significantly different gap structure. This suggests that planets that open similar gap structures, regardless of the cooling timescale, undergo a jump at similar radii.

Nevertheless, we note that $\beta_{\m{fld}}$ only cools locally. The full inclusion of FLD may be relevant, for example because $\beta_{\m{fld}}$ overestimates the vortensity growth rate (\citealt{ziamprasMigrationLowmassPlanets2024}). This should be further investigated in future work.

\subsection{Lifetime of radial structures}\label{sub:lifetimes}

Most of the created dust structures remain visible in the intensity image at $t=500$\,kyr. Our isothermal simulation confirms the results of \citetalias{wafflard-fernandezIntermittentPlanetMigration2020}, \citetalias{mcnallyMigratingSuperEarthsLowviscosity2019}, and \citetalias{meinersPlanetMigrationALMA}. As \citetalias{meinersPlanetMigrationALMA} point out in their analysis of different viscosity levels, long-lasting structures are only present for $\alpha\leq 10^{-4}$.

\citetalias{wafflard-fernandezIntermittentPlanetMigration2020} propose that the lifetime of these structures depends on the time spent in slow migration before or between jumps. To validate this statement in radiative discs, we additionally evaluated the simulations presented in Appendix~\ref{appendix A}. Comparing the pressure maxima, minima, and surface density profiles, we find a clear correlation between the lifetimes of the structures and the duration of slow migration (vortex-assisted migration in this paper), in line with \citetalias{wafflard-fernandezIntermittentPlanetMigration2020}. We find no direct correlation between the cooling timescale and the dissipation of pressure maxima (although we do identify an indirect connection, since the cooling timescale influences the planets' migration and number of jumps). This is expected because the lifetime of these structures is tied to the diffusion (i.e. viscous) timescale, which does not depend on the cooling timescale.

\subsection{Vortex lifetimes}\label{sec: discussion vortices}

We observe large-scale vortices ($\sim$3--5\,au in radial width) in models \texttt{iso} and \texttt{beta}, but not in models \texttt{adapt} or \texttt{adapt-low-mass}. They are decoupled from the planet, do not influence its migration, and persist for an average of $120$\,kyr in mode \texttt{iso} and $55$\,kyr in model \texttt{beta}. This is short compared to the disc's lifetime and the simulation runtime \citep[consistent with][]{rometschSurvivalPlanetinducedVortices2021,rafikov-cimerman-2023}. Thus, dust asymmetries caused by vortices are expected to be less commonly observed than ring-like structures. Additionally, they are only created in simulation with less realistic heating and cooling implementations. 

An additional comparison of vortex lifetimes across the different simulation (including Appendix~\ref{appendix A}) reveals that vortices survive only marginally longer for $\beta = 0.01$ than for $\beta = 1$, and no vortices are created if $\beta >1$. This partially agrees with \cite{rometschSurvivalPlanetinducedVortices2021}, which finds the shortest vortex lifetimes for $\beta = 1$ and longer lifetimes for $\beta < 1$ than for $\beta > 1$. The planet mass used in both studies differs significantly, since \cite{rometschSurvivalPlanetinducedVortices2021} used a Jupiter-mass planet. This may explain parts of the discrepancy. 

We also observe that the locations of the large-scale vortices coincide with the created pressure maxima in models \texttt{iso} and \texttt{beta}. In model \texttt{adapt-low-mass}, the pressure maxima still appear but are considerably weaker, without any pre-existing large-scale vortices. Furthermore, the lifetime of the vortices does not determine the visibility and dissipation of the dust structures. Therefore, contrary to the results of \citetalias{meinersPlanetMigrationALMA}, for discs with more realistic cooling timescales, pressure maxima are not strictly tied to large-scale vortices.

\begin{figure}[t!]
    \includegraphics[width=0.24
    \textwidth]{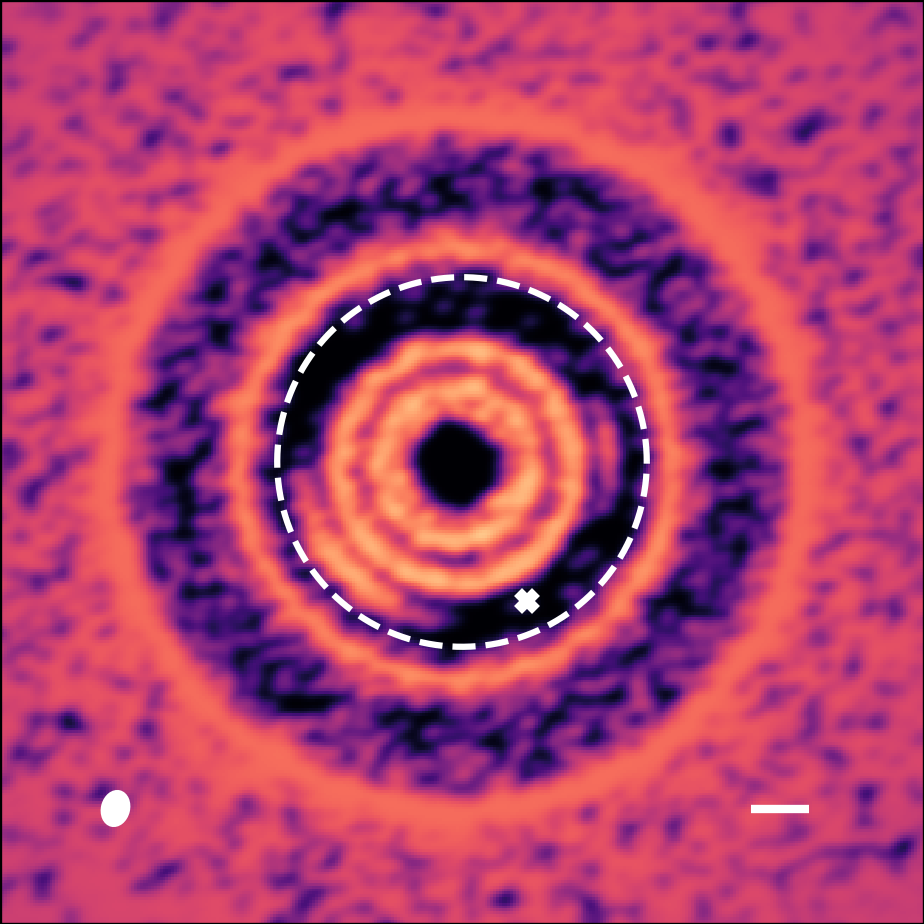}
    \includegraphics[width=0.24\textwidth]{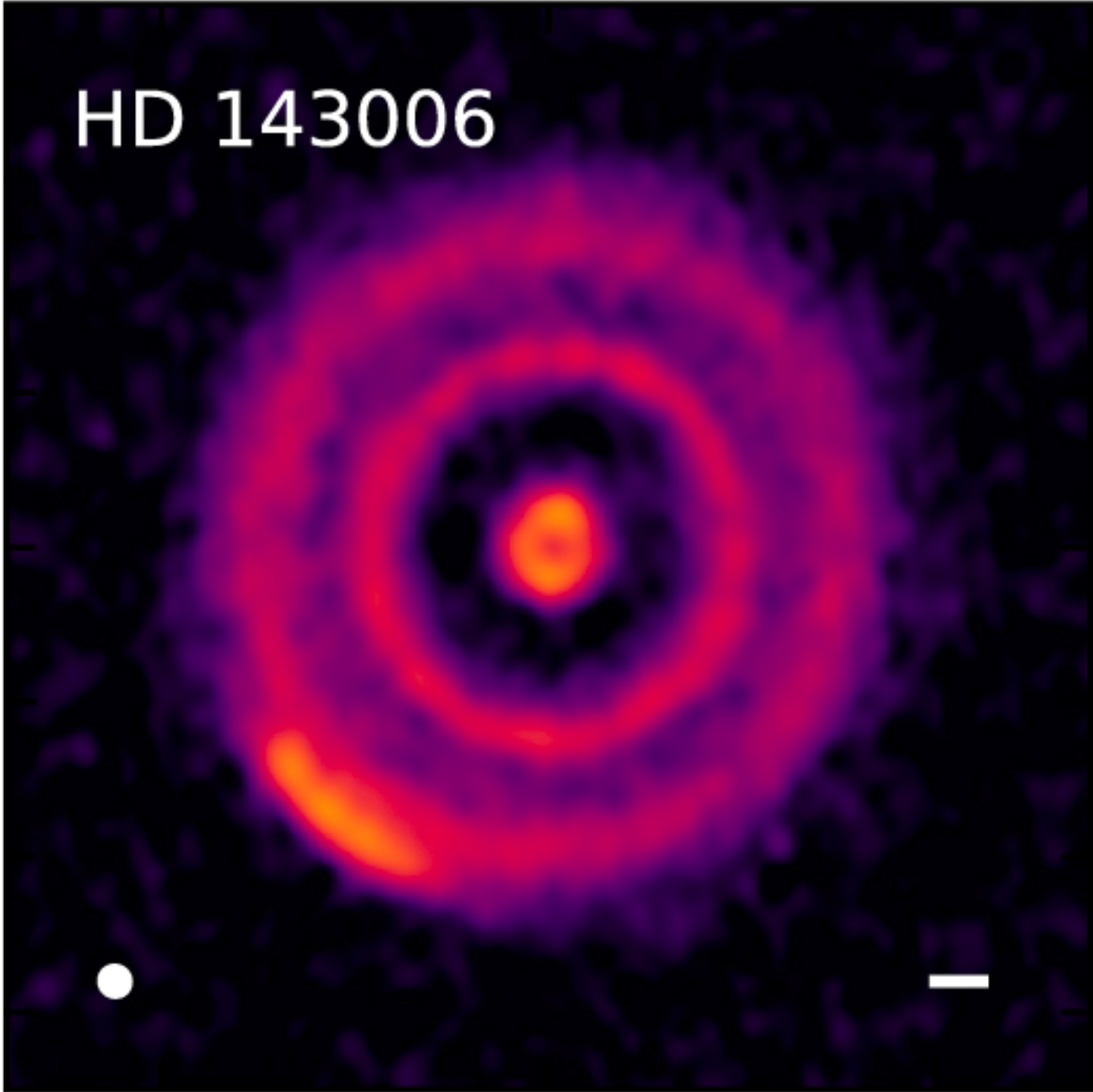}
\caption{Comparison of the convolved intensity image of model \texttt{iso} at t=300\,kyr (left), placed at a distance of 100 pc, and HD~143006 (right). The image of HD~143006 is taken from \citet{andrewsDiskSubstructuresHigh2018}. Both images include the 10\,au scale bar.}
\label{comparison_HD_143006}
\end{figure}

\subsection{Comparison to HD 143006}

We find that the inner disc of model \texttt{iso} resembles the observations of HD~143006 \citep{andrewsDiskSubstructuresHigh2018} reasonably well (Fig.~\ref{comparison_HD_143006}). Both present an asymmetry, which in the case of model \texttt{iso} is created by material at the L5 Lagrange point, in addition to two rings in continuum emission. When omitting the outer disc of model \texttt{iso} and comparing the inner discs, they qualitatively resemble each other (i.e. our model uses a smaller beam, and the colour scale differs between the two panels). However, a locally isothermal model is far from realistic for modelling migration jumps in real discs (see the discussion in Sects.~\ref{sec:const beta} onwards), and the ring expected to exist exterior to the planet's initial location (at $R\approx60$\,au) is missing. Hence, we caution against over-interpreting our results in the context of this system.

\section{Conclusions} \label{sec:conclusion}

This paper demonstrates that one single migrating planet can produce several visible dust rings and gaps for different EOS. These structures remain visible for at least 300--400\,kyr after their formation and likely persist for several hundred thousand years longer, a sizeable fraction of the median disc lifetime (\citealt{armitageAstrophysicsPlanetFormation2020}). Therefore, we can potentially explain the dynamical history of observed discs that show similar structures. We also confirm that the lifetimes of the produced dust structures depend on the time spent in slow migration before or between migration jumps \citepalias[as in][]{wafflard-fernandezIntermittentPlanetMigration2020}, even in radiative discs. 

We also see that cooling substantially affects the migration behaviour of a planet. The resulting asymmetric baroclinic forcing is an additional source of vortensity in the horseshoe region, which may accelerate migration if the planet has not yet opened a deep gap, as seen in models \texttt{beta} and \texttt{adapt}. The time required for a planet to clear its gap at its final location strongly depends on whether it has migrated far enough into the optically thick part of the disc, where cooling can assist in the gap-opening process \citepalias[see][]{ziamprasHaltingMigrationSuperEarths2025}.

Some simulations showed large-scale vortices, always tied to the $\mathcal{IV}$ maxima. With an average vortex lifetime of only $\sim 90$\,kyr in our models, we should not expect protoplanetary discs with asymmetries caused by them to be as common as those without, if they are related to gap-opening planets. They also do not appear in the simulations with the most realistic cooling prescription, likely due to their faster dissipation due to cooling \citep[e.g.][]{fung-ono-2020,rometschSurvivalPlanetinducedVortices2021}. This does not include asymmetries created by other mechanisms, such as material left at the Lagrange point, which is observed in all simulations. 

Our results indicate that migrating planets could be the origin of some of the substructure observed in millimetre emission with ALMA and highlight the importance of including a realistic cooling prescription when modelling this behaviour. As such, care should be taken when modelling and interpreting such structures. 
Differentiating between observable signatures of migrating or non-migrating planets, or of other origins, is beyond the scope of this paper but could be the focus of further work.

\begin{acknowledgements}
K.M.W. would like to thank Anna Penzlin for their advice and very insightful discussions. K.M.W, C.P.D and S.C.M. acknowledge support by the High Performance and Cloud Computing Group at the Zentrum für Datenverarbeitung of the University of Tübingen, the state of Baden-Württemberg through bwHPC and the German Research Foundation (DFG) through grant no INST 37/935-1 FUGG. K.M.W. and A.Z. acknowledge funding from the European Union under the European Union’s Horizon Europe Research and Innovation Programme 101124282 (EARLYBIRD). Views and opinions expressed are, however, those of the authors only and do not necessarily reflect those of the European Union or the European Research Council. Neither the European Union nor the granting authority can be held responsible for them. All plots in this paper were made with the \texttt{python} package \texttt{matplotlib} \citealt{Hunter:2007}).
\end{acknowledgements}

\section*{Data Availability}
Data from our numerical models are available upon reasonable request to the corresponding author.

\bibliographystyle{aa}
\bibliography{references1}

\begin{appendix}

\section{Migration tracks of different $\beta$-cooling simulations}\label{appendix A}

\begin{figure}[h!]
    \centering
    \includegraphics[width=\columnwidth]{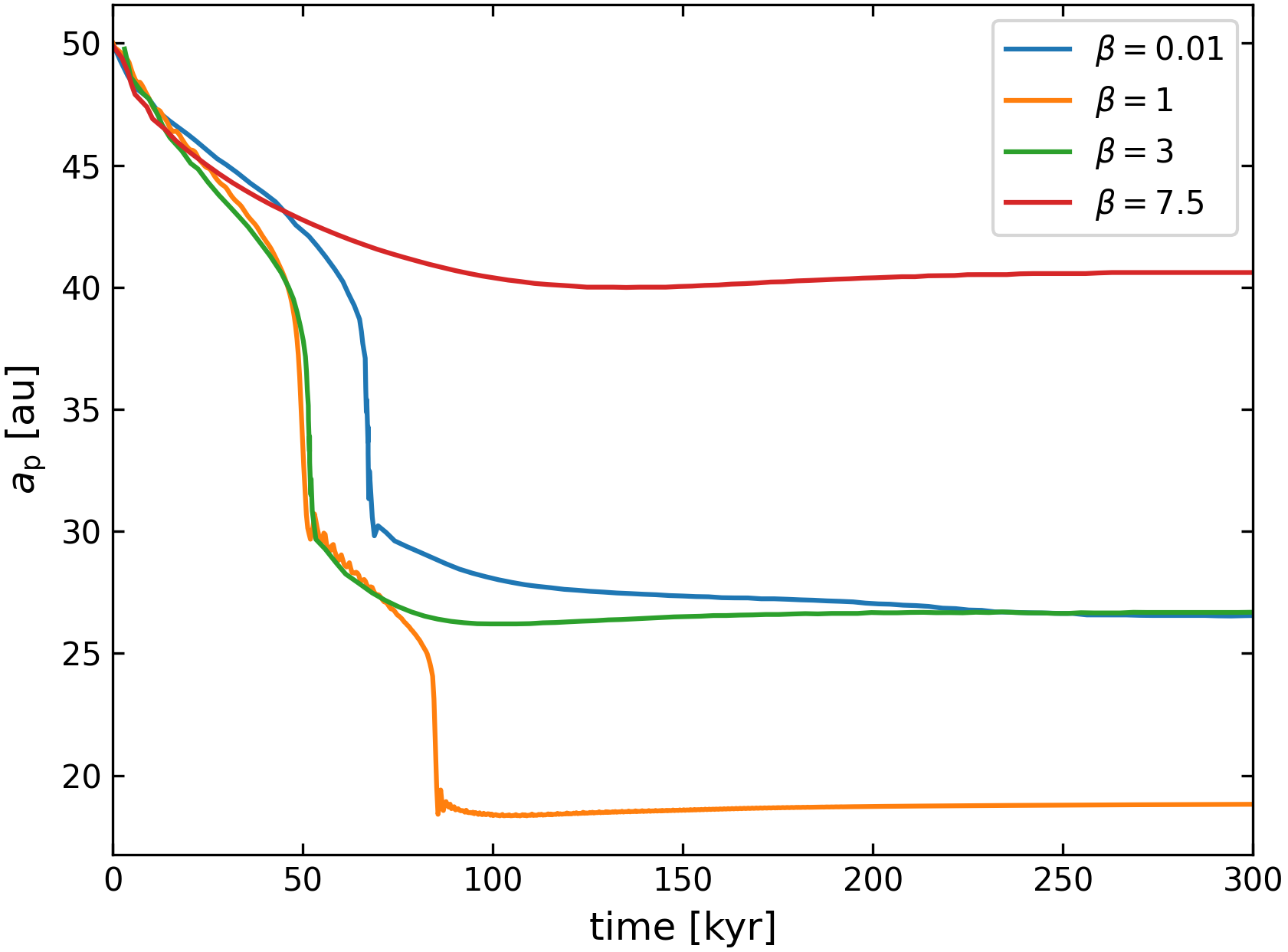}
     \caption{Semi-major axis evolution of model \texttt{beta} compared to simulations with $\beta = 0.01,\, 3$ and $7.5$.}
      \label{fig5ap}
\end{figure}

Figure~\ref{fig5ap} displays the semi-major axis evolution of simulations with different constant $\beta$-cooling timescales, compared to model \texttt{beta}. We ran these additional models to further analyse the radius, where the planet undergoes its first migration jump as well as the vortex and dust structure lifetimes as a function of the cooling timescale for our analysis in Sects.~\ref{discussion:migration}~and~\ref{sec: discussion vortices}. We observe that the planet in the simulations with $\beta = 0.01$ and $3$ undergoes one migration jump at the same radius as model \texttt{beta}. Unsurprisingly, the simulation with $\beta = 0.01$ is very similar to our isothermal model. On the other hand, the planet in the simulation with $\beta = 7.5$ does not undergo a migration jump, since the vortices did not accelerate the planet’s migration sufficiently for it to catch up with its own gap.

\section{Baroclinic forcing and low-vortensity ribbons}
\label{appendix B}

\begin{figure*}
    \centering
    \includegraphics[width=\textwidth]{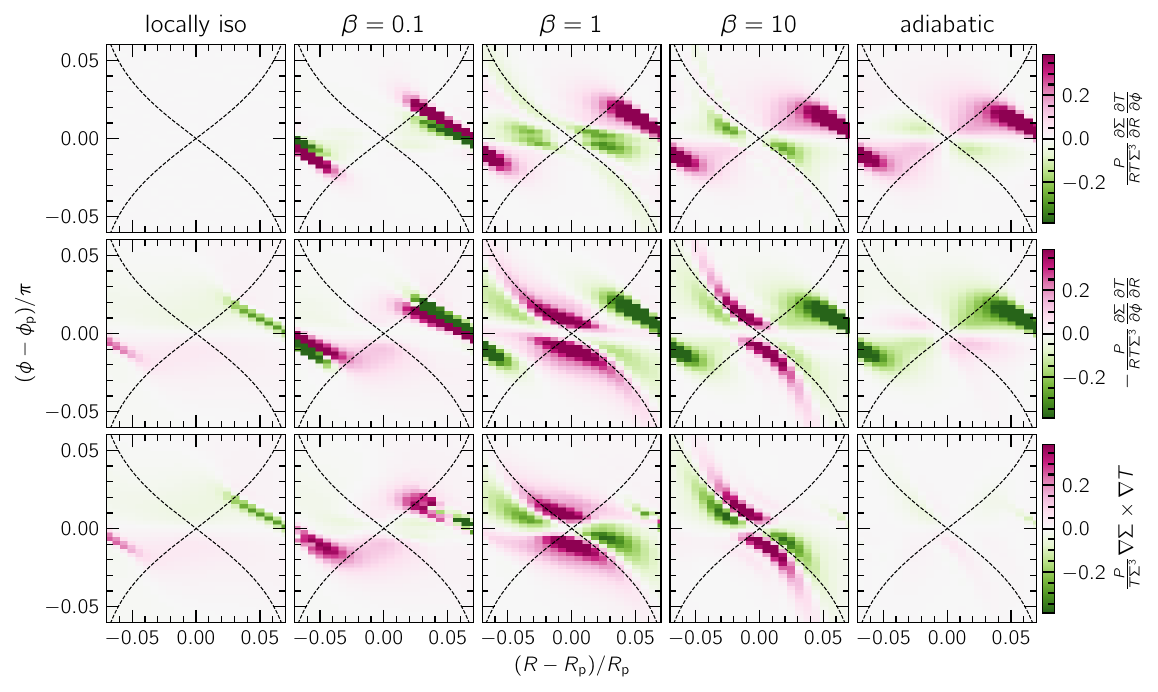}
     \caption{Different components of the baroclinic forcing term as in Eqs.~\eqref{baroclinic forcing}~\&~\eqref{eq:forcing-components} for our \texttt{PLUTO} models with different cooling prescriptions. Dashed curves follow the edges of the horseshoe region, with the planet at the centre of each panel. We can identify the vortensity sourcing within the horseshoe region as in \citet{ziamprasMigrationLowmassPlanets2024} (purple patches at $R\approx R_{\textrm{p}}$) as well as the vortensity sinks associated with the high-$\mathcal{IV}$ ribbons in Fig.~\ref{vortices iso} for $\beta\sim1$ (green patches at $\phi\approx\phi_{\textrm{p}}$).}
      \label{fig:forcing}
\end{figure*}

In Sect.~\ref{sec:const beta} we discussed the mechanism behind the formation of a high-$\mathcal{IV}$ (i.e. low-vortensity) ribbon exterior to the planet's orbit, notably outside of the planet's horseshoe region. As mentioned in the main text, this ribbon is the product of significant baroclinic forcing as gas interacts with the shock front at $\phi\approx\phi_\text{p}$ before shearing past the planet. This suggests that there should be a clear dependence of this forcing term on the temperature gradient across the shock $\partial T/\partial\phi$, and, therefore, on the cooling timescale.

To investigate this, we ran a set of five models using the \texttt{PLUTO} code \citep{mignone-etal-2007,mignone-etal-2012}, mirroring the setup in Sect.~\ref{sec:const beta} but using a lower resolution of $N_R\times N_\phi=256\times768$ and holding the planet fixed at 50\,au for 10 orbits. Each of these five models uses a different cooling prescription: locally isothermal ($\beta=0$), cooling with $\beta\in[0.1, 1, 10]$, and adiabatic ($\beta\rightarrow\infty$). We then compute the individual components of the baroclinic forcing term at $t=10$ planetary orbits as in Eq.~\eqref{baroclinic forcing}:

\begin{equation}
    \label{eq:forcing-components}
    \mathcal{S}_1 = \frac{P}{T R\Sigma^3}\frac{\partial \Sigma}{\partial R} \frac{\partial T}{\partial \phi},\quad\mathcal{S}_2 = -\frac{P}{T R\Sigma^3}\frac{\partial \Sigma}{\partial \phi} \frac{\partial T}{\partial R},
\end{equation}

with $\mathcal{S} = \mathcal{S}_1 + \mathcal{S}_2$. The results are shown in Fig.~\ref{fig:forcing}, which illustrates that this baroclinic forcing does indeed depend strongly on the prescribed cooling timescale. In particular, we find that the forcing responsible for the high-$\mathcal{IV}$ ribbons (identified by green patches close to the planet in the bottom row, as in Fig.~\ref{baroclinic forcing}) peaks for $\beta\sim1$ (at least among the values tested), in line with the discussion in Sect.~\ref{sec:const beta}.

By plotting the individual components of $\mathcal{S}$, we can identify that this ribbon is indeed driven by term $\mathcal{S}_1$ (top row) and is mostly sensitive to $\partial T/\partial\phi$ (see also Fig.~\ref{explanation BF}). Since $\partial T/\partial\phi=0$ in our locally isothermal models, this term is completely absent there, hence the lack of a high-$\mathcal{IV}$ ribbon in the model discussed in Sect.~\ref{sec:iso}. We can further note that the term $\mathcal{S}_2$ (middle row) peaks within the horseshoe region (here marked approximately by black dashed curves), not contributing to this behaviour. Instead, these peaks are related to the mechanism described in \citet{ziamprasMigrationLowmassPlanets2024}, driving vortensity growth within the horseshoe region and contributing to a negative dynamical corotation torque in addition to the vortex-assisted refilling of the gap described in the main text. Ultimately, both processes operate in tandem to accelerate migration inwards, explaining the more vigorous ''jumping'' that the planet experiences in model \texttt{beta}.

\end{appendix}

\end{document}